\documentclass[sn-mathphys-num]{sn-jnl}

\usepackage{graphicx}%
\usepackage{multirow}%
\usepackage{amsmath,amssymb,amsfonts}%
\usepackage{amsthm}%
\usepackage{mathrsfs}%
\usepackage[title]{appendix}%
\usepackage{xcolor}%
\usepackage{textcomp}%
\usepackage{manyfoot}%
\usepackage{booktabs}%
\usepackage{algorithm}%
\usepackage{algorithmicx}%
\usepackage{algpseudocode}%
\usepackage{listings}%
\usepackage{fancyhdr}
\usepackage{natbib}
\usepackage{bm}
\newtheorem{theorem}{Theorem}
\renewcommand{\footnoterule}{%
	
	\hrule width 0.3\textwidth height 0.4pt
	\kern 6pt  
}

\newtheorem{lemma}{Lemma}%

\newtheorem{definition}{Definition}%
\usepackage{hyperref}
\begin{document}
	
	\title[Article Title]{Semigroup-homomorphic Signature}

	\author[1,2]{\fnm{Heng} \sur{Guo}}\email{guoheng@ruc.edu.cn}
	
	\author*[1,3]{\fnm{Kun} \sur{Tian}}\email{tkun19891208@ruc.edu.cn}

	\author[3]{\fnm{Fengxia} \sur{Liu}}\email{shunliliu@gbu.edu.cn}
	
	\author[1,3]{\fnm{Zhiyong} \sur{Zheng}}\email{zhengzy@ruc.edu.cn}
	
	\affil[1]{\orgdiv{School of Mathematics}, \orgname{Renmin University of China}, \orgaddress{ \city{Beijing},  \country{China}}}
	
	\affil[2]{\orgdiv{Institute of Interdisciplinary Science}, \orgname{Renmin University of China}, \orgaddress{ \city{Beijing},  \country{China}}}
	
	\affil[3]{\orgname{Great Bay University}, \orgaddress{ \city{Dongguan}, \state{Guangdong Province}, \country{China}}}

	
	\abstract{In 2002, Johnson et al. posed an open problem at the Cryptographers' Track of the RSA Conference: how to construct a secure homomorphic signature on a semigroup, rather than on a group. In this paper, we introduce, for the first time, a semigroup-homomorphic signature scheme. Under certain conditions, we prove that the security of this scheme is based on the hardness of the Short Integer Solution (SIS) problem and is tightly secure. Furthermore, we extend it to a linear semigroup-homomorphic signature scheme over lattices, and this scheme can also ensure privacy. }
	

	
	
	
	\maketitle

\section{Introduction}
The primary purpose of homomorphic signature schemes was to provide authentication services for network coding and to effectively mitigate pollution attacks\cite{1}. However, the significance of these schemes extends well beyond this initial goal. Due to their capability to perform computations on authenticated data, they have proven to be of unique importance across a variety of application scenarios. The idea of homomorphic signature schemes was initially proposed by R. Rivest \cite{2} during a lecture at Cambridge University in 2000. In 2002, Johnson et al. \cite{3} formalized the definition and conducted a security analysis of these schemes. Following this, a range of different homomorphic signature schemes has been developed. When categorizing these schemes according to the type of homomorphic operation they support, they are typically divided into linearly homomorphic signatures, homomorphic signature scheme for polynomial functions, and fully homomorphic signatures.

The first homomorphic signature scheme to be proposed was the linearly homomorphic signature scheme. In 2007, Zhao et al. put forward a linearly homomorphic signature scheme \cite{1}. This scheme enables arbitrary linear combinations of signed data, which is highly beneficial for verifying the integrity of received messages. Moreover, it plays a crucial role in effectively safeguarding applications based on network coding against pollution attacks. Following this pioneering work, a plethora of efficient and practical linearly homomorphic signature schemes \cite{4,5,6,7,8,9,10} have been introduced. Researchers have since dedicated their efforts to enhancing these schemes, with a particular focus on improving efficiency, bolstering security, and strengthening privacy protection.

In addition to linearly homomorphic signatures, more flexible homomorphic signature schemes for polynomial functions and fully homomorphic signatures have also been developed. In 2011, Boneh et al. proposed the first homomorphic signature scheme for polynomial functions, which allows multivariate polynomial computations on signed data \cite{11}. There have not been many improvements to homomorphic signature schemes for polynomial functions. Currently, only Hiromasa et al. \cite {12}, Arita et al. \cite {13} and Catalano et al. \cite {14} have made meaningful contributions to the efficiency and security of these schemes.  In 2009, Gentry et al. proposed the first fully homomorphic encryption scheme \cite{15}, which attracted widespread attention from cryptographers and subsequently led many researchers to investigate fully homomorphic signatures. In 2014, Gorbunov et al. proposed the first leveled fully homomorphic signature scheme \cite{16}. Subsequently, the research works in \cite{17,18,19,20,21} have made continuous improvements in efficiency, functionality, and security.

According to the definition of homomorphic signature in reference \cite{3}, the message space $\mathcal{M}$ is not required to be a group, nor does the signature algorithm have to be a group homomorphism. Furthermore, in certain group homomorphic algorithms, when the signature of the identity element is known or the signature of the inverse of a specific message can be readily computed, homomorphic operations can be exploited to generate the signature of any message. This situation would render the signature system ineffective. If the message space is a semigroup instead of a group, the signature is then called a semigroup-homomorphic signature, as described in reference \cite{3}. Consequently, in addressing the above-mentioned issue, a semigroup-homomorphic signature is comparatively secure.

In light of this, Johnson et al.\cite{3} posed the following open question at the Cryptographers' Track of the RSA Conference: Is it possible to find a concrete example of a secure semigroup-homomorphic signature rather than a group-homomorphic signature?
To be more specific, let $\mathcal{M}=\{0,1\}^{*}$  be the message space, and define the operation on $\mathcal{M}$ as $\parallel$, which represents the concatenation of bit strings. For any $\mathbf{x},\mathbf{y}\in \{0,1\}^{*}$, $\mathbf{x}\parallel \mathbf{y}\in \mathcal{M}$. It is clear that $\mathcal{M}$ forms a semigroup under $\parallel$. Next, let the signature space be $\Sigma$, with an operation defined on  $\Sigma$ as $\otimes$, and let the signature algorithm be $\textbf{Sign}:\mathcal{M}\rightarrow\Sigma$. The question is whether it is possible to construct a semigroup-homomorphic algorithm \textbf{Sign}, such that the following holds:
\[
\textbf{Sign}(\mathbf{x} \parallel \mathbf{y}) = \textbf{Sign}(\mathbf{x}) \otimes \textbf{Sign}(\mathbf{y}), \quad \text{for any} \ \mathbf{x}, \mathbf{y} \in \{0,1\}^{*}.
\]
However, if we take such a semigroup $\mathcal{M}=\{0,1\}^{*}$ as the message space, there will be a fatal flaw: once the signatures of 0 and 1 are given, then through the homomorphic algorithm, the signatures of any message $x\in\{0,1\}^{*}$ will be obtained, which makes the signature system meaningless. Therefore, it is necessary to consider new semigroups.

\vspace{1\baselineskip}
\textbf{Our Contribution.} In this paper, we propose the first semigroup-homomorphic signature scheme. Under certain conditions, we show that the security of this scheme is based on the hardness of the Short Integer Solution (SIS) problem, and we prove that it is tightly secure. Furthermore, we extend the scheme to a linear semigroup-homomorphic signature scheme based on lattices, which also ensures privacy

\vspace{1\baselineskip}
\textbf{Overview of our techniques.}
\begin{enumerate}
  \item[$\bullet$] \textbf{Determining the Message Space and Signature Space.}

We know that homomorphic mappings preserve algebraic structures. Homomorphic signature algorithms can be regarded as homomorphic mappings between two algebraic structures. Therefore, if the message space forms a semigroup under a certain operation, the signature space will also form at least a semigroup under a corresponding operation.

As mentioned earlier, we cannot choose the semigroup $(\{0,1\}^{*}, \parallel)$; instead, we need to select a different semigroup. If we write $\{0,1\}^{*}$ as $(\mathbb{Z}_{2})^{*}$, then in general, $(\mathbb{Z}_{p})^{*}$ denotes the set of elements where each position is taken from $\mathbb{Z}_{p}$. Let $\mathbb{Z}_{+}=\{0, 1, 2, \dots\}$. Then $(\mathbb{Z}_{+})^{*}$ denotes the set of elements where each position is taken from $\mathbb{Z}_{+}$.

It is evident that $((\mathbb{Z}_{p})^{*}, \parallel)$ and $((\mathbb{Z}_{+})^{*}, \parallel)$ are both semigroups. When $p$ is sufficiently large, if the homomorphic signature uses these two semigroups as the message space, it avoids the situation where the signature of all messages can be derived from the signatures of a few messages through homomorphic operations.

To avoid additional requirements (such as $p$ being sufficiently large), in our construction, we choose $((\mathbb{Z}_{+})^{*}, \parallel)$ as the message space for our homomorphic signature scheme. Finally, we choose $((\mathbb{Z}^{n})^{*}, \parallel)$ as the signature space of the homomorphic signature scheme. Similarly, here $(\mathbb{Z}^{n})^{*}$ denotes the set where each position of the elements comes from $\mathbb{Z}^{n}$. In fact, this is a natural choice.

\vspace{1\baselineskip}
  \item[$\bullet$]  \textbf{Signature generation for the semigroup-homomorphic signature.}

  We primarily leverage the trapdoor generation algorithm \textbf{TrapGen}$(q,h,n)$ of Peikert et al. \cite{22} for key pair generation and the pre-image sampling algorithm \textbf{SamplePre}($A$,$T_{A}$, $\mathbf{u}$, $\sigma$) of  Gentry et al.\cite{23} for signature generation. Let $\mathbf{x}\in ((\mathbb{Z}_{+})^{*},\parallel)$, where  $\mathbf{x}=(\mathbf{x}_{1},...,\mathbf{x}_{k})$ and $\mathbf{x}_{i}\in \mathbb{Z}_{+}$ for $1\leq i\leq k$. To sign the message $\mathbf{x}$,  we perform a sub-signature operation on each component $\mathbf{x}_{i}$ of the message $\mathbf{x}$ to obtain $\bm{\sigma}_{i}$, and then apply a homomorphic operation to  $\bm{\sigma}_{1},..., \bm{\sigma}_{k}$. The resulting signature for the message $\mathbf{x}$ is $\bm{\sigma}=\bm{\sigma}_{1}\parallel\ldots\parallel\bm{\sigma}_{k}$.

Next, we describe how to generate sub-signatures. First, we select a collision-resistant hash function $\mathbf{h}:\mathbb{Z}_{+}=\{0,1\}^{\ast}\rightarrow\{0,1\}^{k}$.
Then, we apply the hash function $\mathbf{h}$ to the component  $\mathbf{x}_{i}$ of the message $\mathbf{x}$, obtaining a $k$-dimensional bit-string $\mathbf{v}_{i}$. Drawing on the idea from \cite{24}, we randomly and independently choose  $k$ linearly independent vectors $\bm{\alpha}_{1},\bm{\alpha}_{2}...,\bm{\alpha}_{k}$ in $\mathbb{Z}_{q}^{h}$(which are part of the public key). Then, we set $\bm{\beta}_{i}=\sum_{j}^{k}\mathbf{v}_{ij}\bm{\alpha}_{j}$.
Finally, by using the sampling algorithm  \textbf{SamplePre}($A$,$T_{A}$, $\bm{\beta}_{i}$, $\sigma$), we can obtain a sub-signature $\bm{\sigma}_{i}$.

\vspace{1\baselineskip}
  \item[$\bullet$] \textbf{Linear operations on semigroup-homomorphic signatures.}

  Based on the definition and security model of the linearly homomorphic signature scheme provided in \cite{11}, we can similarly define the linear semigroup-homomorphic signature scheme and its security model (see Section 2).  In  our linear semigroup-homomorphic signature scheme, to ensure the completeness of linear operations, we define the signature of the empty  string $\varnothing$ as $\aleph$, which satisfies that for any  $\mathbf{x}\in (\mathbb{Z}_{+})^{*}$, $\mathbf{x}\parallel\varnothing=\varnothing\parallel \mathbf{x}=\mathbf{x}$ and for any $\bm{\sigma}\in (\mathbb{Z}^{n})^{*}$, $\bm{\sigma}\parallel \aleph=\aleph\parallel\bm{\sigma}=\bm{\sigma}$.
  Next,  we define $0\cdot \mathbf{x}=\varnothing$ and $0\cdot \bm{\sigma}=\aleph$ for any $\mathbf{x}\in (\mathbb{Z}_{+})^{*}$ , $\bm{\sigma}\in (\mathbb{Z}^{n})^{*}$.  If $c\in \mathbb{Z}_{p}=\{0,1,2,...,p-1\}$ and $c\neq0$, $\mathbf{x}\in( \{0,1\}^{*}, \parallel)$, we define
    \[
c \cdot\mathbf{x} = \underbrace{\mathbf{x} \parallel \cdots \parallel \mathbf{x}}_{c }.
\]
Therefore, if \( \langle f \rangle = (c_1, \dots, c_k) \), where \( c_i \in \mathbb{Z}_p \), then for any \( \mathbf{x}_1, \dots, \mathbf{x}_k \in (\mathbb{Z}_{+})^* \), we have

\[
f(\mathbf{x}_1, \dots, \mathbf{x}_k) = c_1 \cdot\mathbf{x}_1 \parallel \cdots \parallel c_k \cdot\mathbf{x}_k = \underbrace{\mathbf{x}_1 \parallel \cdots \parallel \mathbf{x}_1}_{c_1 } \parallel \cdots \parallel \underbrace{\mathbf{x}_k \parallel \cdots \parallel \mathbf{x}_k}_{c_k }.
\]
Similarly, we can define the linear operations on the signature space $(\mathbb{Z}^{n})^{*}$. For the sake of readability, we will not repeat it here.

\vspace{1\baselineskip}
\item[$\bullet$] \textbf{Signature generation for the linear semigroup-homomorphic signature.}

In essence, the only difference between the linear semigroup-homomorphic signature we constructed and the semigroup-homomorphic signature constructed above is that in the linear semigroup-homomorphic signature, each data set is bound to a unique label vector \(\bm{\tau}\). Therefore, the label vector needs to be properly handled during the signature process. Drawing on the idea from \cite{24}, we map the label vector \(\bm{\tau}\) to a diagonal matrix $H$ whose diagonal elements are $\pm 1 $. Then, by means of the algorithm $\textbf{NewBasis}(A,T_{A},H)$\cite{24}, we generate a new basis $T_{B}$, where $T_{B}$ is the basis of the lattice $\Lambda_{q}^{\bot}(B)$ and  $B=AH^{\top}$.  The advantage of this algorithm is that the dimensions of the newly generated lattice and basis remain unchanged. Finally, the way of generating the signature is consistent with the signature generation process of the above semigroup-homomorphic signature.
\end{enumerate}

\textbf{Paper Organization.} In Section 2, we introduce the definition of semigroup-homomorphic signature and their security model, followed by the definition and security model of linear semigroup-homomorphic signature. In Section 3, we present some commonly used notations and preliminary knowledge. In Section 4, we propose our semigroup-homomorphic signature and prove its correctness and security. In Section 5, we introduce linear semigroup-homomorphic signature and prove their correctness, unforgeability, and privacy. The final section concludes the paper and discusses some open problems.

\section{ Semigroup-Homomorphic Signature: Definition and Security model}

\subsection{Semigroup-Homomorphic Signature}

A signature scheme is defined by a message space $\mathcal{M}$, a set of private keys $\mathcal{K}$, a set of public keys $\mathcal{K}^{\prime}$, a signature generation algorithm $\textbf{Sign}: \mathcal{K} \times \mathcal{M} \rightarrow \mathcal{Y}$ (which may be randomized), and a verification algorithm $\textbf{Verify}: \mathcal{K}^{\prime} \times \mathcal{M} \times \mathcal{Y} \rightarrow \{0, 1\}$. The scheme must satisfy the condition that for all messages $x \in \mathcal{M}$, when $(k, k^{\prime})$ are matching private and public keys, it holds that $\textbf{Verify}(k^{\prime}, x, \textbf{Sign}(k, x)) = 1$. We often omit the private and public keys, using $\textbf{Sign}(x)$ in place of $\textbf{Sign}(k, x)$ and $\textbf{Verify}(x, s)$ instead of $\textbf{Verify}(k^{\prime}, x, s)$, provided this causes no ambiguity. Additionally, for a binary operation $\odot: \mathcal{M} \times \mathcal{M} \rightarrow \mathcal{M}$ and a subset $S \subseteq \mathcal{M}$, we define $\mathrm{span}_{\odot}(S)$ as the smallest set $T$ such that $S \subseteq T$ and $x \odot y \in T$ for all $x, y \in T$.

In 2002, Johnson et al.\cite{3} first formally defined homomorphic signatures and the associated security model. Below, based on \cite{3}, we provide the original definition of homomorphic signature.

\vspace{1\baselineskip}
\begin{definition}\label{d1}(homomorphic signature,\cite{3})
  Let $\textbf{Sign}: \mathcal{K}\times \mathcal{M}\rightarrow \mathcal{Y}$ and  $\textbf{Verify}: \mathcal{K}^{\prime}\times \mathcal{M}\times \mathcal{Y}\rightarrow 1$ be a signature scheme, and let $\odot: \mathcal{M }\times\mathcal{M }\rightarrow \mathcal{M}$ be a binary operation. We say that  $\textbf{Sign}$ is homomorphic with respect to $\odot$ if there exists an efficient family of binary operations $\otimes_{k^{\prime}}: \mathcal{Y}\times\mathcal{Y}\rightarrow \mathcal{Y}$  such that for all $x,x^{\prime}, y, y^{\prime}$ satisfying
   $\textbf{Verify}(x,y)=\textbf{Verify}(x^{\prime},y^{\prime})=1,$  the following condition holds:
  $$y\otimes_{k^{\prime}}y^{\prime}=\textbf{Sign}(k, x\odot x^{\prime}).$$
\end{definition}

\vspace{1\baselineskip}
A homomorphic signature scheme is called a semigroup-homomorphic signature scheme when the message space forms a semigroup, not a group. Below, we provide the definition of a semigroup-homomorphic signature.

\vspace{1\baselineskip}
\begin{definition}\label{d2}
In the description of the homomorphic signature mentioned above, if the message space $(\mathcal{M},\odot)$ is a semigroup rather than a group, then we refer to this homomorphic signature as a semigroup-homomorphic signature.
\end{definition}

\vspace{1\baselineskip}
For the homomorphic signature scheme we construct, we introduce a definition of existential unforgeability under the adaptive chosen-message attack with a fixed message range (EUF-CMA-FMR), which can be viewed as a selective version of the existential unforgeability definition introduced in \cite{3}.

We introduce a weaker security concept named EUF-CMA-FMR because there are some fundamental obstacles in achieving the standard concept in \cite{3}. The obstacles are as follows: According to the definition of our signature algorithm, in order to avoid trivial forgeries, we require that only messages $\mathbf{x}\in \mathbb{Z}_{+}$ can be signed with the private key. Therefore, there is a certain range limitation on the messages for signature queries. This motivates the introduction of our security concept, EUF-CMA-FMR, in this paper.

\vspace{1\baselineskip}
\begin{definition}\label{d3}(EUF-CMA-FMR)
We say that a homomorphic signature scheme $\mathcal{S}$ is $(t,q,\epsilon)$-secure  under the adaptive chosen-message attacks  within a fixed message range , if each adversary, after being given the public key, makes at most $q$ adaptive chosen-message queries to a non-empty subset $\mathcal{M}^{\prime}$ of the message space $\mathcal{M}$, and runs in time at most $t$,  with its advantage satisfying $\mathrm{Adv}(\mathcal{A})\leq \epsilon$. The advantage of an adversary $\mathcal{A}$ is defined as the probability that,
after queries on the messages $x_{1},...,x_{q}\in \mathcal{M}^{\prime},$ $\mathcal{A}$ outputs a valid signature $(x^{\prime},y^{\prime})$ on
some message $x^{\prime}\notin \mathrm{span}_{\odot}(x_{1},...,x_{q})$.  In other words,
$$\mathrm{Adv}(\mathcal{A})=\mathrm{Pr }[\mathcal{A}^{\textbf{Sign}(k,\cdot)}=(x^{\prime},y^{\prime})\wedge \textbf{Verify}(x^{\prime},y^{\prime})=1\wedge x^{\prime}\notin \mathrm{span}_{\odot}(x_{1},...,x_{q}) ].  $$
\end{definition}

\textbf{Remark 1}: To ensure that the subset $\mathcal{M}^{\prime}$ has enough messages to provide for the adversary's queries, it is generally required that the number of elements in $\mathcal{M}^{\prime}$ is much greater than the maximum number of queries, i.e.,  $|\mathcal{M}^{\prime}|\gg q$

\subsection{Linear Semigroup-Homomorphic Signature}

Inspired by the definition of linearly homomorphic signature schemes in \cite{5,11}, we present the following definition of a linear semigroup-homomorphic signature scheme.

\vspace{1\baselineskip}
\begin{definition}\label{d6.1}(Linear semigroup-homomorphic signature )
 A linearly semigroup-homomorphic signature scheme  is a tuple of probabilistic, polynomial-time algorithms $\mathcal{LHS}=$\textbf{(Setup, Sign, Combine, Verify)} with the following functionality:

\vspace{1\baselineskip}
\begin{itemize}
  \item[$\bullet$]  \textbf{Setup}$(n, pp)$. Given a security parameter $n$  and additional public parameters $pp$
 , the public parameter $pp$ defines a maximum data-set size $k$, a message space $(\mathcal{M}, \oplus)$,  and a signature space $(\Sigma,\otimes)$, where the message space is a semigroup,  not a group.  The algorithm then outputs a public key $pk$ and a secret key $sk$.

\vspace{1\baselineskip}
\item[$\bullet$]  \textbf{Sign}$(sk, \tau, \vec{\mathbf{v}})$. Given a secret key $sk$, data-set tag $\tau\in\{0,1\}^{n}$,  and a vector  $\vec{\mathbf{v}}\in \mathcal{M}$,  this algorithm outputs a signature $\bm{\sigma}$.

\vspace{1\baselineskip}
  \item[$\bullet$]  \textbf{Combine}$(pk,\tau, \{(c_{i}, \bm{\sigma}_{i})_{i=1}^{l}\})$.  Given a public key $pk$, a data-set tag $\tau\in\{0,1\}^{n}$, and a set of tuples $(c_{i}, \bm{\sigma}_{i})_{i=1}^{l}$ where $c_{i}\in R$ and $\bm{\sigma}_{i}\leftarrow \textbf{Sign}(sk, \tau, \vec{\mathbf{v}_{i}})$, with $R$ being an integer ring, this algorithm outputs a signature $\sigma$, which is meant to be a signature on $\sum_{i=1}^{l}c_{i}\vec{\mathbf{v}}_{i}$.

 \vspace{1\baselineskip}
  \item[$\bullet$] \textbf{Verify}$(pk, \tau, \vec{\mathbf{y} }, \bm{\sigma})$.  On input a public key $pk$, an data-set tag $\tau\in\{0,1\}^{n}$, a vector $\vec{\mathbf{y}}\in \mathcal{M}$, and  a signature $\bm{\sigma}$,  this algorithm outputs either 0 (reject) or 1 (accept).

\end{itemize}

We require that for each $(pk,sk)$ output by \textbf{Setup}$(n, pp)$, we have:
\begin{enumerate}
  \item [(1)] For all $\tau$ and $\vec{\mathbf{y}}\in \mathcal{M}$, if $\bm{\sigma}\leftarrow \textbf{Sign}(sk, \tau, \vec{\mathbf{y}})$  then $\textbf{Verify}(pk, \tau, \vec{\mathbf{y}} , \sigma)=1.$

  \vspace{1\baselineskip}
  \item [(2)] For all $\tau\in \{0.1\}^{n}$  and all sets of triples $\{(c_{i}, \bm{\sigma}_{i}, \vec{\mathbf{v}}_{i})\}_{i=1}^{l}$, if $\textbf{Verify}(pk, \tau, \vec{\mathbf{v}}_{i} , \bm{\sigma}_{i})=1$, for all $i$, then
      $$  \textbf{Verify}(pk, \tau, \sum_{i=1}^{l}c_{i}\vec{\mathbf{v}}_{i}, \textbf{Combine}(pk,\tau, \{(c_{i}, \bm{\sigma}_{i})\}_{i=1}^{l}) ) =1. $$

\end{enumerate}
\end{definition}

\vspace{1\baselineskip}
Since our linear semigroup-homomorphic signature scheme is a direct extension of the semigroup homomorphism signature scheme constructed in Section 4, we will similarly present the definition of unforgeability under adaptive chosen - message attacks within a fixed message range for the linear semigroup - homomorphic signature, abbreviated as Unforgeability-FMR.

\vspace{1\baselineskip}
\begin{definition}\label{d6.2}(Unforgeability-FMR)
A linearly semigroup-homomorphic signature scheme $\mathcal{LHS}$ = (\textbf{Setup},\textbf{ Sign}, \textbf{Combine}, \textbf{Verify})
is unforgeable if the advantage of any probabilistic, polynomial-time adversary $\mathcal{A}$ in the following
security game is negligible in the security parameter $n$:

\begin{enumerate}
\item[] \textbf{Setup:} The challenger runs \textbf{Setup}$(n, pp)$ to obtain $(pk,sk)$, and gives $pk$ to $\mathcal{A}$.

\item[] \textbf{Queries}: The challenger selects a sufficiently large subset $\mathcal{M}^{\prime} \subseteq \mathcal{M}$, where $|\mathcal{M}^{\prime}|\gg qk$. Proceeding adaptively, $\mathcal{A}$ specifies a sequence of subsets
\[
V_{i} \subseteq \mathcal{M}^{k}, \quad V_{i} = \operatorname{span}_{\oplus}\{\mathbf{v}_{i1}, \mathbf{v}_{i2}, \dots, \mathbf{v}_{ik}\}, \quad \mathbf{v}_{ij} \in \mathcal{M}^{\prime},
\]
where $1 \leq i \leq q$, and $1 \leq j \leq k$.  For each $i$, the challenger chooses a tag $\tau_{i}$ uniformly from $\{0,1\}^{n}$ and gives to $\mathcal{A}$ the pair $(\tau_{i}, \{\bm{\sigma}_{ij}\})$, where the signatures are computed as:
\[
\bm{\sigma}_{ij} \leftarrow \textbf{Sign}(sk, \tau_{i}, \mathbf{v}_{ij}), \quad j = 1, \dots, k.
\]

\item[] \textbf{Output}: $\mathcal{A}$ outputs $\tau^{*}\in\{0,1\}^{n}$, $\mathbf{y}^{*}\in \mathcal{M}$, and a signature $\bm{\sigma}^{*}$
\end{enumerate}

The adversary wins if \textbf{Verify}$(sk,\tau^{*},\vec{\mathbf{y}}^{*}, \bm{\sigma}^{*})=1$, and either

\begin{itemize}
  \item [(1)] $\tau^{*}\neq \tau_{i}$ for all $i$ (\textbf{ type I forgery}), or
  \item [(2)] $\tau^{*}= \tau_{i}$ for some $i$ but $\vec{\mathbf{y}}^{*}\notin V_{i}$ ( \textbf{type II forgery}).
\end{itemize}
The advantage of $\mathcal{A}$ is defined to be the probability that $\mathcal{A}$  wins the security game.
\end{definition}

\vspace{1\baselineskip}
The following is a review of the definition of weakly context hiding \cite{11}

\vspace{1\baselineskip}
\begin{definition}\label{d6}(Weakly context hiding)
A  linear semigroup-homomorphic signature scheme $\mathcal{LHS}=$(\textbf{Setup}, \textbf{Sign}, \textbf{Combine},\textbf{ Verify}) is weakly
context hiding  if the advantage of any probabilistic, polynomial-time adversary $\mathcal{A}$ in the following security
game is negligible in the security parameter $n$:

\vspace{1\baselineskip}
\textbf{Setup}: The challenger runs \textbf{Setup}$(n, pp)$ to obtain $(pk,sk)$ and gives $pk$ and $sk$ to $\mathcal{A}.$

\vspace{1\baselineskip}
\textbf{Challenge}: A outputs $(V_{0}, V_{1}, f_{1},..., f_{s})$, where $V_{b}=span_{\oplus}\{\mathbf{v}_{b1},..., \mathbf{v}_{bk}\}$, $\{\mathbf{v}_{b1},..., \mathbf{v}_{bk}\}\subseteq \mathcal{M}$, $b=0,1.$  The functions $f_{1},...,f_{s}$ are R-linear functions from $\mathcal{M}^{k}$ to $\mathcal{M}$, and they satisfy
$$  f_{i}(\mathbf{v}_{01},..., \mathbf{v}_{0k})=f_{i}(\mathbf{v}_{11},...,\mathbf{v}_{1k}),  1\leq i\leq s  .$$
In response, the challenger generates a random bit $b\in\{0,1\}$ and a random tag $\tau\in\{0,1\}^{n}$ and signs the
vector subsets $V_{b}$ using the tag $\tau$. Next, for $i=1,...,s$, the challenger uses the algorithm \textbf{Combine} to derive signatures $\bm{\sigma}_{i}$ on $f_{i}(\mathbf{v}_{b1},...,\mathbf{v}_{bk})$  and sends $\bm{\sigma}_{1},..., \bm{\sigma}_{s}$ to $\mathcal{A}$

\vspace{1\baselineskip}
\textbf{Output}: $\mathcal{A}$ outputs a bit $b^{\prime}.$

The adversary $\mathcal{A}$ wins the game if $b=b^{\prime}.$ The advantage of $\mathcal{A}$ is the probability that $\mathcal{A}$ wins the game.
\end{definition}

\subsection{Tight security}

At the end of this section, we provide a brief introduction to the important concept of tight security in cryptography.

In the framework of provable security, reduction is a key concept for evaluating the security of signature schemes. The definition is as follows. 

\vspace{1\baselineskip}
\begin{definition}\label{d6.2222}
If an adversary $\mathcal{A}$ breaks the scheme $\mathcal{S}$ with $(t, \epsilon)$ in the defined security model, then there exists an algorithm $\mathcal{B}$ that breaks a certain computational problem $P$ with $(t', \epsilon')$, where $\epsilon' = \epsilon/\theta$ and $t' = t + o(t)$, and $\theta \geq 1$.The parameter $\theta$ is used to measure the tightness of the reduction. In this context, the security parameter is denoted as $\lambda$, and the number of adversarial queries is $Q$. When $\theta = \mathcal{O}(1)$, it is called a tight reduction; when $\theta = \text{poly}(\lambda)$, it is an almost tight reduction; when $\theta = \text{poly}(Q)$, it is a loose reduction.
\end{definition}

\vspace{1\baselineskip}
The use of tight reductions is of great significance. From a practical perspective, a tighter reduction allows for shorter security parameters, thereby improving efficiency. Theoretically, a tight reduction indicates that the difficulty of the two computational problems is close. 

\section{Primitives}

\textbf{Notation}.  In our discussion, we employ the notations $\mathcal{O}$, $\widetilde{\mathcal{O}}$, and $\omega$ to characterize the growth rates of functions. Consider two functions $f$ and $g$ that depend on the variable $n$. We say that the relationship $f(n) = \mathcal{O}(g(n))$ is valid precisely when there exist a positive constant $c$ and an integer $N$ such that for every integer $n$ greater than $N$, the inequality $f(n) \leq c \cdot g(n)$ is satisfied. 

Similarly, we define the notation $f(n) = \widetilde{\mathcal{O}}(g(n))$ to represent a scenario where, for a specific positive constant $c'$, the function $f(n)$ satisfies the relation $f(n) = \mathcal{O}(g(n) \cdot \log^{c'} n)$. As for the notation $f(n) = \omega(g(n))$, it is established if and only if there is an integer $N$ such that for every positive constant $c$ and all integers $n$ that are greater than $N$, the inequality $g(n) \leq c \cdot f(n)$ holds.

Let the security parameter be denoted as $n$. When there exists a positive constant $c$ such that $f(n)=\mathcal{O}(n^c)$, we write $f(n)=\mathrm{poly}(n)$. On the contrary, if for every positive constant $c$, the function $f(n)$ has the property that $f(n)=\mathcal{O}(n^{-c})$, then $f(n)$ is referred to as negligible and is symbolized as $\mathrm{negl}(n)$. In the case where the probability of an event taking place is $1-\mathrm{negl}(n)$, we state that the event occurs with overwhelming probability.

We use the capital letter $\mathbb{Z}$ to denote the ring of all integers. Meanwhile, $\mathbb{Z}_q$ stands for the ring of integers modulo $q$ (where $q \geq 2$), and $\mathbb{R}^n$ represents the $n$-dimensional Euclidean space. In our notation, we employ capital letters like $A, B, C$, etc., to represent matrices, and bold lowercase letters such as $\mathbf{a}, \mathbf{b}, \mathbf{c}$, etc., to denote vectors. 

Suppose $A = (\mathbf{a}_1, \dots, \mathbf{a}_n) \in \mathbb{R}^{h \times n}$. The norm of matrix $A$ is defined in the following way: $\|A\| = \max_{1 \leq i \leq n} \|\mathbf{a}_i\|$, where $\|\mathbf{a}_i\|$ represents the $l_2$-norm of the vector $\mathbf{a}_i$.

Moreover, we denote $\widetilde{A} = (\widetilde{\mathbf{a}}_1, \dots, \widetilde{\mathbf{a}}_n)$ as the result of applying the Gram - Schmidt orthogonalization process to matrix $A$. The Gram-Schmidt orthogonalization is defined as follows:
\[
\widetilde{\mathbf{a}}_1 = \mathbf{a}_1, \quad
\widetilde{\mathbf{a}}_i = \mathbf{a}_i - \sum_{j = 1}^{i - 1} \frac{\langle \mathbf{a}_i, \widetilde{\mathbf{a}}_j \rangle}{\langle \widetilde{\mathbf{a}}_j, \widetilde{\mathbf{a}}_j \rangle} \widetilde{\mathbf{a}}_j, \quad 2 \leq i \leq n,
\]
where $\langle \cdot, \cdot \rangle$ denotes the standard inner product within the Euclidean space.

For an arbitrary distribution $\mathcal{D}$, the notation $x\sim \mathcal{D}$ indicates that the variable $x$ is distributed according to $\mathcal{D}$, and $x\leftarrow \mathcal{D}$ represents the operation of sampling a random value in accordance with this distribution. Given a set $\mathcal{X}$, the expression $x \stackrel{\$}{\leftarrow} \mathcal{X}$ is used to signify the act of uniformly and randomly selecting an element $x$ from $\mathcal{X}$. For any probabilistic polynomial-time (PPT) algorithm $Alg$, we use $y\leftarrow Alg(x)$ to convey that the algorithm takes $x$ as input and subsequently yields the output $y$.

\vspace{1\baselineskip}
\begin{definition}\label{d3.1}(\emph{Lattice}\cite{24})
Let $\Lambda\subset \mathbb{R}^{n}$ be a non-empty subset. $\Lambda$ is called a lattice if:

(1) it is an additive subgroup of $\mathbb{R}^{n}$;

(2) there exists a positive constant $\lambda=\lambda(\Lambda)>0$ such that
$$\mathrm{ min}\{\parallel \mathbf{x}\parallel | \mathbf{x}\in\Lambda , \mathbf{x}\neq 0 \}=\lambda.$$  $\lambda$ is called the minimum distance.
\end{definition}
A full-rank $n$-dimensional lattice can also be expressed as a linear combination of a set of basis vectors $B=\{\mathbf{b}_{1},...,\mathbf{b}_{n}\}\subset \mathbb{R}^{n}$:
$$ \Lambda=\mathcal{L}(B)=\{ B\cdot \mathbf{x}=\sum_{i=1}^{n}x_{i}\mathbf{b}_{i}| x=(x_{1},...,x_{n})^{\top}\in \mathbb{Z}^{n}\}.$$
We call $\Lambda^{*}$  the dual lattice of $\Lambda$ if $\Lambda^{*}=\{\mathbf{y}\in \mathbb{R}^{n}| <\mathbf{y},\mathbf{x}> \in \mathbb{Z} \quad\text{for all } \mathbf{x}\in\Lambda \}$.

\vspace{1\baselineskip}
\begin{definition}\label{d3.2}(\emph{$q$-ary lattices}, \cite{26})
Let $A\in \mathbb{Z}_{q}^{n\times m}$, $\mathbf{u}\in \mathbb{Z}^{n}$. The following two $q$-ary lattices are defined as:

(1)$\Lambda_{q}^{\bot}=\{ \mathbf{x}\in \mathbb{Z}^{m} | A\cdot \mathbf{x} \equiv 0 (\mathrm{mod} q)   \}$

(2)$\Lambda_{q}^{\mathbf{u}}=\{ \mathbf{y}\in \mathbb{Z}^{m} | A\cdot \mathbf{y} \equiv \mathbf{u} (\mathrm{mod} q)   \}$

The set $\Lambda_{q}^{\mathbf{u}}$ is a coset of  $\Lambda_{q}^{\bot}$ since $\Lambda_{q}^{\mathbf{u}} =\Lambda_{q}^{\bot}+\mathbf{t}$ for any $\mathbf{t}$ such that $A\cdot \mathbf{t}\equiv \mathbf{u} (\mathrm{mod}q)$.
\end{definition}

\vspace{1\baselineskip}
The short integer solution (SIS) problem is formally described as follows.

\vspace{1\baselineskip}
\begin{definition}\label{d3.3}(\emph{Short integer solution} \cite{27})
Let $n$, $m$, $q$ be positive integers, with $m=\mathrm{poly(}n)$. Let $A\in \mathbb{Z}_{q}^{n\times m}$ be a uniformly distributed random matrix over $\mathbb{Z}_{q}$, and let $\beta\in \mathbb{R}$ such that $0<\beta<q$. The SIS problem is to find a short integer solution $\mathbf{x}$ satisfying the following condition:
$$ A\cdot \mathbf{x}\equiv 0(\mathrm{mod}q), \quad \mathrm{and}\quad \mathbf{x}\neq0, \parallel \mathbf{x}\parallel\leq\beta.$$
We write the above SIS problem as $\mathrm{SIS}_{q,n,m,\beta}$ or $\mathrm{SIS}_{q,\beta}$.
\end{definition}

\vspace{1\baselineskip}
Ajtai, in \cite{28}, established that solving the SIS problem is at least as hard as solving a related worst-case problem. Later, Gentry et al., in \cite{23}, presented an improved reduction, detailed as follows.

\vspace{1\baselineskip}
\begin{theorem}\label{t3.1}(\emph{Worst-case to average-case reduction }\cite{23})
For any polynomial bounded \(m = \text{poly}(n)\), and any \(\beta > 0\), if \(q \geq \beta \cdot \omega(\sqrt{n\log n})\), then solving the average-case problem \(\textbf{SIS}_{q,\beta}\) is at least as hard as solving the worst-case problem \(SIVP_{\gamma}\) on any \(n\)-dimensional lattice for \(\gamma=\beta \cdot \tilde{O}(\sqrt{n})\).

\end{theorem}

\vspace{1\baselineskip}
\begin{definition}\label{d3.4}( \emph{Gaussian distributions}\cite{27})
Let $s$ be a positive real number and $\mathbf{c}\in \mathbb{R}^{n}$ be a vector. The Gaussian function centered at $\mathbf{c}$ with parameter $s$ is defined as: $\rho_{s,\mathbf{c}}(\mathbf{x})=e^{\frac{-\pi}{s^{2}}\parallel\mathbf{x}-\mathbf{c}\parallel^{2}}.$ The discrete Gaussian measure $\mathcal{D}_{\Lambda,s, \mathbf{c}}$ defined on the lattice $\Lambda$ is given by:
$$  \mathcal{D}_{\Lambda,s, \mathbf{c}}=\frac{\rho_{s,\mathbf{c}}(\mathbf{x})}{\rho_{s,\mathbf{c}}(\Lambda)},$$
where $\rho_{s,\mathbf{c}}(\Lambda)=\sum_{\mathbf{x}\in \Lambda}\rho_{s,\mathbf{c}}(x)$.
\end{definition}

\vspace{1\baselineskip}
\begin{definition}\label{d3.4}( \emph{Statistical distance}\cite{27})
Let $M\subset\mathbb{R}^{n}$ be a finite or countable set, and let $X$ and $Y$ be two discrete random variables taking values in $M$. The statistical distance between $X$ and $Y$ is defined as:
$$ \bigtriangleup (X,Y)=\frac{1}{2}\sum_{a\in M}|P\{X=a\}-P\{Y=a\}|.$$
\end{definition}
When the statistical distance between two distributions is less than a negligible value, we say that the two distributions are statistically  close.

\vspace{1\baselineskip}
\begin{definition}\label{d3.44}( \emph{Min-entropy}\cite{16})
For a random variable \( X \), its min-entropy is defined as:
\[
H_{\infty}(X) = -\log\left( \max_{x \in X} \Pr[X = x] \right).
\]

The average conditional min-entropy of a random variable \( X \) conditional on a correlated variable \( Y \) is defined as:
\[
H_{\infty}(X|Y) = -\log\left( \mathbb{E}_{y \in Y} \left\{ \max_{x \in X} \Pr[X = x | Y = y] \right\} \right).
\]
\end{definition}

\vspace{1\baselineskip}
The optimal probability of an unbounded adversary guessing $X$ given the correlated value $Y$ is $2^{-H_{\infty}(X|Y)}$.

\vspace{1\baselineskip}
\begin{lemma}\label{l3.1}(\cite{16})
Let $X$, $Y$ be arbitrarily random variables where the support of $Y$ lies in  $\mathcal{Y}$, Then
$$ H_{\infty}(X|Y)> H_{\infty}(X)-\log(|\mathcal{Y}|). $$
\end{lemma}

\vspace{1\baselineskip}
Micciancio and Goldwasser \cite{29} proved that a full-rank set $S$ in the lattice $\Lambda$ can be transformed into a basis $T$ such that both have similarly low Gram-Schmidt norms.

\vspace{1\baselineskip}
\begin{lemma}\label{l3.1}(\cite{29}, \emph{Lemma 7.1})
Let $\Lambda$ be an $n$-dimensional lattice. There is a deterministic  polynomial-time algorithm that, given an arbitrary basis of and a full-rank set $S=\{s_{1},...,s_{n}\}$ in $\Lambda$, returns a basis $T$ of $\Lambda$ satisfying
$$ \parallel \widetilde{T}\parallel\leq\parallel \widetilde{S}\parallel \text{and} \parallel T\parallel \leq \parallel S\parallel \frac{\sqrt{n}}{2}. $$
\end{lemma}

\vspace{1\baselineskip}
Ajtai \cite{30}, and later Alwen and Peikert \cite{22}, presented a method for sampling a matrix $A \in \mathbb{Z}_{q}^{n \times m}$ that is statistically close to uniform, along with an associated basis $S_A$ for the lattice $\Lambda_{q}^{\perp}(A)$, which has a low Gram-Schmidt norm. The result below is derived from Theorem 3.2 in \cite{22}, where the parameter $\delta$ is set to $1/3$. The theorem guarantees the construction of a matrix $A$ that is statistically close to uniformly distributed in  $\mathbb{Z}_{q}^{n \times m}$, accompanied by a short basis. Given that $m$ is significantly larger than $n$, the matrix $A$ is rank $n$ with overwhelming probability. Therefore, the theorem can be interpreted as stating that $A$ is statistically close to a uniform rank $n$ matrix from  $\mathbb{Z}_{q}^{n \times m}$.

\vspace{1\baselineskip}
\begin{theorem}\label{t3.2}(\cite{22})
Let $q\geq3$ be odd, $n$ be a positive integer,  and let $m:=\lceil 6n\log q\rceil.$  There is a probabilistic
polynomial-time algorithm \textbf{TrapGen}$(q,n,m)$ that outputs a pair $(A\in \mathbb{Z}_{q}^{h\times n}, T_{A}\in \mathbb{Z}^{n\times n})$ such that $A$ is statistically close to a uniform rank $n$ matrix in $\mathbb{Z}_{q}^{n\times m}$, and $T_{A}$ is a basis for $\Lambda_{q}^{\bot}(A)$ satisfying
$$ \parallel\widetilde{T_{A}}\parallel\leq \mathcal{O}(\sqrt{n\log q}) \quad\text{and}\quad \parallel T_{A}\parallel \leq \mathcal{O}(n\log q).$$
\end{theorem}

\vspace{1\baselineskip}
Gentry et al.\cite{23} propose an algorithm capable of sampling from a discrete Gaussian distribution utilizing an arbitrary short basis.

\vspace{1\baselineskip}
\begin{lemma}\label{l3.3}(\emph{Sampling from discrete Gaussian }\cite{23})
Let $q\geq2$ , $A\in \mathbb{Z}_{q}^{n\times m}$ with $m>n$ and let $T_{A}$ be a basis for $\Lambda_{q}^{\bot}(A)$ and $s\geq \widetilde{T_{A}}\cdot\omega(\sqrt{\log m })$. Then for $\mathbf{c}\in \mathbb{R}^{n}$ and $\mathbf{u}\in \mathbb{Z}_{q}^{n}$:
\begin{enumerate}
  \item  There is a  probabilistic polynomial-time algorithm \textbf{SampleGaussian}$(A,T_{A},s,\mathbf{c})$ that outputs $\mathbf{x}\in \Lambda_{q}^{\bot}(A)$ drawn from a distribution statistically close to $\mathcal{D}_{\Lambda_{q}^{\bot}(A),s,\mathbf{c}}$
  \item  There is a  probabilistic polynomial-time algorithm \textbf{SamplePre}$(A,T_{A},\mathbf{u},s)$ that outputs  $\mathbf{x}\in \Lambda_{q}^{\mathbf{u}}(A)$  sampled from a distribution statistically close to  $\mathcal{D}_{\Lambda_{q}^{\mathbf{u}}(A),s}$, whenever $\Lambda_{q}^{\mathbf{u}}(A)$ is not empty.
\end{enumerate}
\end{lemma}

When $s\geq\omega(\sqrt{\log n})$, we denote the Gaussian sampling algorithm over the integer lattice $\mathbb{Z}^{n}$ as $\textbf{SampleDom}(1^{n},s)$.  That is, when $\mathbf{x}\stackrel{\$}{\leftarrow}\textbf{SampleDom}(1^{n},s)$, $\mathbf{x}$ is statistically close to the distribution $\mathcal{D}_{\mathbb{Z}^{n},s}$.  And these preimages have a conditional min-entropy of $\omega(\log n)$\cite{23}.

\vspace{1\baselineskip}
\begin{lemma}\label{l3.4}(\cite{23})
Let $n$ and $h$ be positive integers, and $q$  be a prime, such that  $n\geq2h\lg q$. Then for all but a $2q^{-h}$  fraction of all $A\in \mathbb{Z}_{q}^{h\times n}$ and for any $s\geq\omega(\sqrt{\log n})$, the distribution of the $\mathbf{\alpha }=A\cdot \mathbf{x }(\mathrm{mod}q)$
is statistically close to uniform over $\mathbb{Z}_{q}^{h}$, where $\mathbf{x}\stackrel{\$}{\leftarrow}\textbf{SampleDom}(1^{n},s)$.
\end{lemma}

\vspace{1\baselineskip}
\begin{definition}\label{d3.5}( \emph{Smoothing parameter}\cite{31})
For any $n$-dimensional lattice $\Lambda$ and any given $\epsilon>0$, the smoothing parameter of the lattice is defined as
\begin{equation*}
\eta_{\epsilon}(\Lambda)=\min\left\{s > 0 \mid \rho_{\frac{1}{s}}(\Lambda^{*}) < 1 + \epsilon\right\}.
\end{equation*}
\end{definition}

For the vast majority of matrices $A\in \mathbb{Z}_{q}^{h\times n}$, there exists a negligible value \(\epsilon\) such that the smoothing parameter  
\(\eta_{\epsilon}(\Lambda_q^{\perp}(\mathbf{A}))\) is smaller than 
\(\omega(\sqrt{\log n})\):

\vspace{1\baselineskip}
\begin{lemma}\label{l3.5}(\cite{23})
Let $q\geq3$, $h$ and $n$ be positive integers satisfying $n\geq 2h\lg q$. Then there exists a negligible function $\epsilon(n)$ such that
$\eta_{\epsilon}(\Lambda_{q}^{\bot}(A))<\omega(\sqrt{\log n})$ for all but at most a $q^{-h}$ fraction of $A$ in the $\mathbb{Z}_{q}^{h\times n}$ .
\end{lemma}

\vspace{1\baselineskip}
\begin{theorem}\label{t3.33}(\cite{32})
$\boldsymbol{t}_i \in \mathbb{Z}^m$ and $\boldsymbol{x}_i$ are mutually independent random variables sampled from a Gaussian distribution $D_{\boldsymbol{t}_i + \Lambda,\sigma}$ over $\boldsymbol{t}_i + \Lambda$ for $i = 1,2,\cdots,k$ in which $\Lambda$ is a lattice and $\sigma \in \mathbb{R}$ is a parameter. Let $\boldsymbol{c} = (c_1,\cdots,c_k) \in \mathbb{Z}^k$ and $g = \mathrm{gcd}(c_1,\cdots,c_k)$, $\boldsymbol{t} = \sum_{i = 1}^{k} c_i \boldsymbol{t}_i$. If $\sigma > \|\boldsymbol{c}\| \eta_{\epsilon}(\Lambda)$ for some negligible number $\epsilon$, then $\boldsymbol{z}= \sum_{i = 1}^{k} c_i \boldsymbol{x}_i$ statistically closes to $D_{\boldsymbol{t} + g\Lambda,\|\boldsymbol{c}\|\sigma}$.
\end{theorem}

\vspace{1\baselineskip}
The subsequent lemma demonstrates that the lengths of vectors obtained through sampling from a discrete Gaussian distribution are predominantly concentrated within a specific bound.

\vspace{1\baselineskip}
\begin{lemma}\label{l3.2}(\cite{31})
Let $\Lambda$ be an $n$-dimensional lattice, and $T$ be a basis of the lattice $\Lambda$. If $s\geq \parallel \widetilde{T}\parallel\cdot \omega(\sqrt{\log n})$, then for any $\mathbf{c}\in \mathbb{R}^{n}$, we have:
$$ \mathrm{Pr}\{\parallel \mathbf{x}-\mathbf{c}\parallel>s\sqrt{n}:\mathbf{x}{\leftarrow} \mathcal{D}_{\Lambda,s,\mathbf{c}}\}\leq \mathrm{negl}(n) .$$
\end{lemma}

\vspace{1\baselineskip}
Agrawal et al.\cite{22} proposed a new lattice basis delegation algorithm, 
\textbf{BasisDel}$(A,R,T_{A},s)$, in which the lattice dimension remains unchanged during the delegation process. Inspired by their \textbf{BasisDel} algorithm,  Guo et al.\cite{24} developed a new algorithm \textbf{NewBasis}$(A,H,T_{A})$, for generating trapdoors in new lattices. This algorithm is a special case of \textbf{BasisDel}$(A,R,T_{A},s)$ algorithm, where the matrix  $R$ is replaced by an orthogonal matrix $H$.  The detailed conclusions are as follows.

\vspace{1\baselineskip}
\begin{theorem}\label{l3.8} (\cite{24})
Let $q\geq3$ be odd, $h$ be a positive integer, and let $n:=\lceil 6h\log q\rceil.$  There exists a deterministic polynomial-time algorithm denoted as \textbf{NewBasis}$(A, H, T_{A} )$ Its inputs are: a matrix $A\in \mathbb{Z}_{q}^{h\times n}$ of rank $h$, a orthogonal matrix $H\in \mathbb{Z}^{n\times n}$, and $T_{A}$ which is a basis of $\Lambda_{q}^{\bot}(A)$. The output of this algorithm is a basis $T_{B}$ of the lattice $\Lambda_{q}^{\bot}(B)$, where $B=AH^{\top}$, and $\parallel \widetilde{T}_{B}\parallel< \parallel \widetilde{T}_{A}\parallel$, $\parallel T_{B}\parallel< \parallel T_{A}\parallel \sqrt{n}/2$
\end{theorem}

\vspace{1\baselineskip}
The following three lemmas establish fundamental properties of orthogonal matrices and Gaussian distributions over lattices.

\vspace{1\baselineskip}
\begin{lemma}\label{l3.7}(\cite{24})
Let $A\in \mathbb{R}^{n\times n}$ be an $n$-dimensional full-rank matrix, and $H\in \mathbb{R}^{n\times n}$ be an orthogonal matrix. Then,
$$ \parallel HA\parallel=\parallel A\parallel \quad\text{and}\quad   \parallel \widetilde{HA}\parallel=\parallel\widetilde{A}\parallel.$$
\end{lemma}

\vspace{1\baselineskip}
\begin{lemma}\label{l3.7777}(\cite{24})
Let  $H$  be an orthogonal matrix over $\mathbb{Z}^{n\times n}$, that is, $HH^{\top} = I_{n}$. If $ \mathbf{x}\sim \mathcal{D}_{\mathbb{Z}^{n}, \sigma}$,   then $H\mathbf{x} \sim \mathcal{D}_{\mathbb{Z}^{n}, \sigma}$.
\end{lemma}

\vspace{1\baselineskip}
\begin{lemma}\label{l3.777}(\cite{24})
Let \(A \in \mathbb{Z}_q^{n\times m}, s > 0, \mathbf{u} \in \mathbb{Z}^{m}\), and \(\Lambda = \Lambda_q^{\boldsymbol{u}}(A)\). If \(\mathbf{x}\) is sampled from \(\mathcal{D}_{\mathbb{Z}^m, s}\) conditioned on \(A\mathbf{x} \equiv \boldsymbol{u} \pmod{q}\), then the conditional distribution of \(\mathbf{x}\) is \(\mathcal{D}_{\Lambda,s}\).
\end{lemma}

\section{Semigroup homomorphic signature}

\subsection{Construction}

Before presenting our semigroup homomorphic signature, we first need to define our message space and signature space. As introduced in the Introduction section, we choose $\mathcal{M}=((\mathbb{Z}_{+})^{*},\parallel)$ as the message space and $\Sigma=((\mathbb{Z}^{n})^{*},\parallel)$ as the signature space, both of which are semigroups without inverses. Note: if $\bm{\sigma}=\bm{\sigma}_{1}\bm{\sigma}_{2}\cdot\cdot\cdot\bm{\sigma}_{k}\in (\mathbb{Z}^{n})^{*}$, we also treat $\bm{\sigma}$ as a matrix $[\bm{\sigma}_{1},\bm{\sigma}_{2},...,\bm{\sigma}_{k}]\in \mathbb{Z}^{n\times k}$.  If $\mathbf{x}\in (\mathbb{Z}_{+})^{*}$, without loss of generality, assume $\mathbf{x}=\mathbf{x}_{1}\mathbf{x}_{2}\cdot\cdot\cdot \mathbf{x}_{k}\in (\mathbb{Z}_{+})^{*}$. We define $|\mathbf{x}|$ as the length of $\mathbf{x}$, that is, $|\mathbf{x}|=k$.

\vspace{1\baselineskip}
Our semigroup-homomorphic signature scheme,  $\mathcal{SH}=( \textbf{Gen}, \textbf{Sign}, \textbf{Verify})$ works as follows:

\vspace{1\baselineskip}
\textbf{Gen}($1^{n}$): Input the secure parameters $n$. Let $q=\mathrm{poly}(n)$, $k=\mathrm{poly}(n)$  and $q\geq (kn)^{2}$,  $h=\lfloor \frac{n}{6\log q}\rfloor $, and $V=k\sqrt{2n\log q}\cdot\log n$ .  The algorithm generates a pair of public key and secret key as follows:

\begin{enumerate}
  \item [(1)] Compute $(A, T_{A})\longleftarrow \textbf{TrapGen}(q,h ,n)$,  where $A\in \mathbb{Z}_{q}^{h\times n}$  and $T_{A}$ is a basis for the lattice $\Lambda_{q}^{\bot}(A)$.
  \item [(2)] Select a secure collision-resistant hash function  $\mathbf{h}: \mathbb{Z}_{+}=\{0,1\}^{\ast}\rightarrow \{0,1\}^{k}$.  Note that the hash function in our context is regarded as a deterministic algorithm rather than an idealized random oracle.

  \item [(3)] Sample $\bm{\alpha}_1, \ldots, \bm{\alpha}_k \stackrel{\$}{\leftarrow} \mathbb{Z}_q^h$.  We require that $\bm{\alpha}_1, \ldots, \bm{\alpha}_k$ be linearly independent. If not, resampling should be carried out, and this process can be easily accomplished within polynomial time.
  \item [(4)] Output the public key $pk=( A, \mathbf{h},\bm{\alpha}_1, \ldots, \bm{\alpha}_k)$ and the secret key $sk=T_{A}$.
\end{enumerate}

\vspace{1\baselineskip}
\textbf{Sign}$(sk, \mathbf{x})$: Input the secret key $sk=T_{A}$,  the  message $\mathbf{x}=\mathbf{x}_{1},...,\mathbf{x}_{|\mathbf{x}|}\in(\mathbb{Z}_{+})^{*}$. the signing algorithm proceeds as follows:

\begin{enumerate}
  \item [(1)] Compute $\mathbf{v}_{i}=\mathbf{h}(\mathbf{x}_{i})=(\mathbf{v}_{i1},...,\mathbf{v}_{ik})$ and  $\bm{\beta}_{i}=\sum_{j=1}^{k}\mathbf{v}_{ij}\cdot\bm{\alpha}_{j}(\mathrm{mod }q)$,  for $i=1,...,|\mathbf{x}|$

  \item [(2)] Compute $\bm{\sigma}_{i}\leftarrow \textbf{SamplePre}(A, T_{A}, \bm{\beta}_{i}, V)$,  for $i=1,...,|\mathbf{x}|$.
\end{enumerate}
Output signature $\bm{\sigma}=\bm{\sigma}_{1}\bm{\sigma}_{2}\cdot\cdot\cdot\bm{\sigma}_{|\mathbf{x}|}\in (\mathbb{Z}^{n})^{*}$.

\vspace{1\baselineskip}
\textbf{Verify}$(pk,\mathbf{x}, \bm{\sigma})$:  Input the public key  $pk=( A, \mathbf{h},\bm{\alpha}_1, \ldots, \bm{\alpha}_k)$,  $\mathbf{x}=\mathbf{x}_{1}\mathbf{x}_{2}\cdot\cdot\cdot \mathbf{x}_{|\mathbf{x}|}\in \mathcal{M}$ and $\bm{\sigma}$.   The verification process is as follows:
\begin{enumerate}
  \item [(1)]Compute   $\mathbf{v}_{i}=\mathbf{h}(\mathbf{x}_{i})$, 
   $\bm{\beta}_{i}=\sum_{j=1}^{k}\mathbf{v}_{ij}\cdot\bm{\alpha}_{j}(\mathrm{mod }q)$, for $i=1,..,|\mathbf{x}|.$ Let $$B=[\bm{\beta}_{1},..., \bm{\beta}_{|\mathbf{x}|}] \quad\text{and}\quad \bm{\sigma}=\bm{\sigma}_{1}\bm{\sigma}_{2}\cdot\cdot\cdot\bm{\sigma}_{|\mathbf{x}|}=
      [\bm{\sigma}_{1},\bm{\sigma}_{2},...,\bm{\sigma}_{|\mathbf{x}|}].$$
  \item [(2)] Output 1 if all of the following two conditions are satisfied; otherwise, output 0.
  \begin{enumerate}
    \item [1)]  $\|\bm{\sigma}\|\leq  V\cdot \sqrt{kn}$; ($\parallel\cdot\parallel$  represents the matrix norm.)
    \item [2)] $A\cdot\bm{\sigma}(\mathrm{mod} q)=B$. ( Note: $A\cdot\bm{\sigma}(\mathrm{mod} q)=[A\cdot\bm{\sigma}_{1},..., A\cdot\bm{\sigma}_{\mathbf{|x|}}](\mathrm{mod} q)$)
  \end{enumerate}
\end{enumerate}

\vspace{1\baselineskip}
\textbf{Remark 2:}
In our signature scheme, in order to avoid trivial forgeries, we actually sign only the messages $\mathbf{x}_{1},..., \mathbf{x}_{k}\in \mathbb{Z}_{+}$ with the private key, while other messages  $\mathbf{x}\in(\mathbb{Z}_{+})^{*}$, which are not signed with the private key, are derived through homomorphic operations, if $\mathbf{x}\in \mathrm{span}_{\parallel}\{\mathbf{x}_{1},...,\mathbf{x}_{k}\}.$

\subsection{Correctness}

We will now prove that the signature scheme we constructed is correct.

\vspace{1\baselineskip}
\begin{theorem}\label{t0}
The above signature scheme $\mathcal{HS}$ satisfies correctness with overwhelming probability.
\end{theorem}

\begin{proof}
Assume that $\bm{\sigma}=\bm{\sigma}_{1}\bm{\sigma}_{2}\cdot\cdot\cdot\bm{\sigma}_{|\mathbf{x}|}\in (\mathbb{Z}^{n})^{*}$ is the signature of the message $\mathbf{x}=\mathbf{x}_{1}\mathbf{x}_{2}\cdot\cdot\cdot \mathbf{x}_{|\mathbf{x}|}\in(\mathbb{Z_{+}})^{*}$, i.e.,
\[
\bm{\sigma}_{i} \leftarrow \textbf{SamplePre}(A, T_{A}, \bm{\beta}_{i}, V), \quad \text{where} \quad \bm{\beta}_{i}=\sum_{j=1}^{k}\mathbf{v}_{ij}\cdot\bm{\alpha}_{j}(\mathrm{mod }q), \quad 1 \leq i \leq |\mathbf{x}|.
\]

From Theorem \ref{t3.2}, it follows that $\parallel \widetilde{T}_{A}\parallel\leq \mathcal{O}(\sqrt{h\log q})$ with overwhelming probability. Since $V=\sqrt{2n\log q}\cdot\log n$, it follows that
 $$\frac{V}{\parallel\widetilde{T}\parallel}\geq k\cdot\sqrt{\frac{2n}{h}}\cdot \log n \geq \sqrt{\log n},$$
which implies  $V\geq \parallel \widetilde{T}\parallel\cdot \omega(\sqrt{\log n})$.
Therefore, by Lemma \ref{l3.2}, we have $\parallel \bm{\sigma}_{i}\parallel\leq V\cdot \sqrt{n}$ with overwhelming probability.
Thus,
$$ \parallel \bm{\sigma} \parallel = \max\limits_{1 \leq i \leq |x|} \parallel \bm{\sigma}_i \parallel \leq V \cdot \sqrt{n}\leq  V \cdot\sqrt{k\cdot  n} . $$

Finally, according to Lemma \ref{l3.3} we have $A\cdot\bm{\sigma}_{i}(\mathrm{mod} q)=\bm{\beta}_{i},$ for $i=1,...,|\mathbf{x}|.$  Thus
$$ A\cdot\bm{\sigma}(\mathrm{mod} q)=[A\cdot\bm{\sigma}_{1},..., A\cdot\bm{\sigma}_{\mathbf{|x|}}](\mathrm{mod} q)=[\bm{\beta}_{1},..., \bm{\beta}_{|\mathbf{x}|}]=B$$
This completes the proof of the theorem.
\end{proof}

\subsection{Homomorphism}

We will now prove that the signature scheme satisfies homomorphism.

\vspace{1\baselineskip}
\begin{theorem}\label{t1}
The above signature scheme $\mathcal{HS}$ is homomorphic with respect to the operation \( \parallel \). That is, for any \( \mathbf{x}, \mathbf{y} \in (\mathbb{Z}_{+})^{*} \), we have

\[
\text{Sign}(pk, \mathbf{x} \parallel \mathbf{y}) =  \text{Sign}(pk, \mathbf{x}) \parallel \text{Sign}(pk, \mathbf{y}).
\]

\end{theorem}

\begin{proof}
For any \( \mathbf{x}, \mathbf{y} \in (\mathbb{Z}_{+})^{*} \), assume that

\[
\left\{
\begin{array}{l}
\bm{\sigma}_{1} = \bm{\sigma}_{11} \bm{\sigma}_{12} \cdots \bm{\sigma}_{1|\mathbf{x}|} \leftarrow \text{Sign}(pk, \mathbf{x}) \\
\bm{\sigma}_{2} = \bm{\sigma}_{21} \bm{\sigma}_{22} \cdots \bm{\sigma}_{2|\mathbf{y}|} \leftarrow \text{Sign}(pk, \mathbf{y}),
\end{array}
\right.
\]

where

\[
\left\{
\begin{aligned}
    \bm{\sigma}_{1i} &\leftarrow \textbf{SamplePre}(A, T_{A}, \bm{\beta}_{1i}, V), \quad \bm{\beta}_{1i} = \sum_{l=1}^{k}\mathbf{h}(\mathbf{x}_i)_{l}\cdot\bm{\alpha}_{l}(\mathrm{mod} q), \quad i = 1, \dots, |\mathbf{x}|, \\
    \bm{\sigma}_{2j} &\leftarrow \textbf{SamplePre}(A, T_{A}, \bm{\beta}_{2j}, V), \quad \bm{\beta}_{2j} =\sum_{l=1}^{k} \mathbf{h}(\mathbf{y}_j)_{l}\cdot\bm{\alpha}_{l}(\mathrm{mod} q), \quad j = 1, \dots, |\mathbf{y}|.
\end{aligned}
\right.
\]

From the verification algorithm, we know that:
\begin{center}
  $\left\{\begin{array}{l}
A\cdot\bm{\sigma}_{1i}\mathrm{(mod} q)=\bm{\beta}_{1i} , i=1,...,|\mathbf{x}|, \\
A\cdot\bm{\sigma}_{2i}\mathrm{(mod} q)=\bm{\beta}_{2i} , i=1,...,|\mathbf{y}|.
\end{array}\right.$
\end{center}

Let \( B_1 = [\bm{\beta}_{11}, \dots, \bm{\beta}_{1|\mathbf{x}|}] \), \( B_2 = [\bm{\beta}_{21}, \dots, \bm{\beta}_{2|\mathbf{y}|}] \), \( \mathbf{z} = \mathbf{x} \parallel \mathbf{y} \), and \( \bm{\sigma} = \bm{\sigma}_1 \parallel \bm{\sigma}_2 \).
Next, calculate \( \bm{\beta}_i = \sum_{l=1}^{k}\mathbf{h}(\mathbf{z}_i)_{l} \cdot \bm{\alpha}_{l} (\mathrm{mod}\, q) \), for \( 1 \leq i \leq |\mathbf{z}| = |\mathbf{x}| + |\mathbf{y}| \).
Let \( B = [\bm{\beta}_1, \dots, \bm{\beta}_{|\mathbf{z}|}] \), then \( B = [B_1, B_2] \).

Since
\begin{center}
  $\left\{\begin{array}{l}
A\cdot\bm{\sigma}_{1}\mathrm{(mod} q)=B_{1} , \\
A\cdot\bm{\sigma}_{2}\mathrm{(mod} q)=B_{2} .
\end{array}\right.$
\end{center}
Thus, $A\cdot\bm{\sigma}=A\cdot[\bm{\sigma}_{1},\bm{\sigma}_{2}](\mathrm{mod}q)=[B_{1},B_{2}]$.  Clearly, $\parallel\bm{\sigma}\parallel< V\cdot\sqrt{k \cdot n}$.

Therefore, according to the verification algorithm,
$$ \mathrm{\textbf{Verify}}(pk, \mathbf{z}, \bm{\sigma})=\mathrm{\textbf{Verify}}(pk, \mathbf{x}\parallel \mathbf{y}, \bm{\sigma}_{1}\parallel\bm{\sigma_{2}})=1. $$
Therefore,
\[
\text{Sign}(pk, \mathbf{x} \parallel \mathbf{y}) =  \text{Sign}(pk, \mathbf{x}) \parallel \text{Sign}(pk, \mathbf{y}).
\]
\end{proof}

\subsection{Security}

In this section, we will prove that our semigroup-homomorphic signature scheme satisfies EUF-CMA-FMR and is tightly secure. Before presenting the main conclusion, we need to rule out a possible forgery.

\vspace{1\baselineskip}
Let $\mathbf{h}: \mathbb{Z}_{+}\rightarrow \{0,1\}^{k}$ be a collision-resistant hash function. If one can find $\mathbf{x}_{1},\mathbf{x}_{2}\in \mathbb{Z}_{+}$ with $\mathbf{x}_{1}\neq \mathbf{x}_{2} $ such that
$\sum_{j=1}^{k}\mathbf{h}(\mathbf{x}_{1})_{j}\cdot\bm{\alpha}_{j}=\sum_{j=1}^{k}\mathbf{h}(\mathbf{x}_{2})_{j}\cdot\bm{\alpha}_{j}(\mathrm{mod }q)$.  According to the signature algorithm, if the signature of $\mathbf{x}_{1}$ is $\bm{\sigma}$, then the signature of $\mathbf{x}_{2}$ will also be $\bm{\sigma}$ thereby resulting in a forgery. However, such a situation does not exist, and we have the following lemma to prove this.

 \vspace{1\baselineskip}
\begin{lemma}\label{sdsd}
Let $\mathbf{h}: \mathbb{Z}_{+}\rightarrow \{0,1\}^{k}$ be a collision-resistant hash function.  If for any $\mathbf{x}_{1},\mathbf{x}_{2}\in \mathbb{Z}_{+}$,  it holds that  $\sum_{j=1}^{k}\mathbf{h}(\mathbf{x}_{1})_{j}\cdot\bm{\alpha}_{j}=\sum_{j=1}^{k}\mathbf{h}(\mathbf{x}_{2})_{j}\cdot\bm{\alpha}_{j}(\mathrm{mod }q)$,  then with overwhelming probability,  $\mathbf{x}_{1}=\mathbf{x}_{2}$.
\end{lemma}

 \vspace{1\baselineskip}
\begin{proof}
According to the definition of the signature algorithm, $\bm{\alpha}_{1},...,\bm{\alpha}_{k}$  are linearly independent. Suppose there exist $\mathbf{x}_{1},\mathbf{x}_{2}\in \mathbb{Z}_{+}$  such that 
 $$\sum_{j=1}^{k}\mathbf{h}(\mathbf{x}_{1})_{j}\cdot\bm{\alpha}_{j}=\sum_{j=1}^{k}\mathbf{h}(\mathbf{x}_{2})_{j}\cdot\bm{\alpha}_{j}(\mathrm{mod }q).$$  
Then it must be the case that  $\mathbf{h}(\mathbf{x}_{1})_{j}=\mathbf{h}(\mathbf{x}_{2})_{j}$ for $j=1,..,k$,  that is,  $\mathbf{h}(\mathbf{x}_{1})=\mathbf{h}(\mathbf{x}_{2})$. Since the hash function $\mathbf{h}$  is collision-resistant, with overwhelming probability, $\mathbf{x}_{1}=\mathbf{x}_{2}$.

\end{proof}

 \vspace{1\baselineskip}
According to Lemma \ref{sdsd}, if a polynomial-time adversary $\mathcal{A}$ finds a pair of distinct messages $(\mathbf{x}_{1},\mathbf{x}_{2})$ such that
 $\sum_{j=1}^{k}\mathbf{h}(\mathbf{x}_{1})_{j}\cdot\bm{\alpha}_{j}=\sum_{j=1}^{k}\mathbf{h}(\mathbf{x}_{2})_{j}\cdot\bm{\alpha}_{j}(\mathrm{mod }q)$,  then it must be the case that  $\mathbf{h}(\mathbf{x}_{1})=\mathbf{h}(\mathbf{x}_{2})$. That is, the adversary $\mathcal{A}$ has found a pair of collision points of the hash function. Obviously, the probability of such a situation is negligible.

\vspace{1\baselineskip}
We now provide the security proof of the proposed signature scheme.

\vspace{1\baselineskip}
\begin{theorem}\label{t2} Suppose $q=\mathrm{poly}(n)$, $k=\mathrm{poly}(n)$  and $q\geq (kn)^{2}$, $h=\lfloor \frac{n}{6\log q}\rfloor$,  $V=k\sqrt{2n\log q}\cdot\log n$. If the $\textbf{SIS}_{q,\beta}$ problem is computationally hard for $\beta=2V\cdot\sqrt{kn}$, then the above semigroup-homomorphic signature $\mathcal{HS}$ is existentially unforgeable under the adaptive chosen-message attack with a fixed message range (EUF-CMA-FMR).

 \vspace{1\baselineskip}
 More specifically,  Suppose the adversary $\mathcal{A}$ is allowed to make at most $q_{s}$ chosen-message queries on message subset $\mathbb{Z}_{+}(\subseteq \mathcal{M}=(\mathbb{Z}_{+})^{*})$ and can break the above semigroup-homomorphic signature scheme $\mathcal{SH}$ with advantage $\epsilon$ within time $t$. Then, the simulator $\mathcal{C}$ can use the adversary's ability to construct an algorithm $\mathcal{B}$ that outputs a solution to the $\textbf{SIS}_{q,2V\cdot\sqrt{kn}}$ problem with advantage $\epsilon-\mathrm{negl}(n)$ within time $t+\mathcal{O}(q_{s}T_{\mathcal{SH}}+T_{\mathrm{SampleDom}})$, where $T_{\mathcal{SH}}$ represents the time required for the simulator to generate a signature, and $T_{\mathrm{SampleDom}}$ represents the time for running the sampling algorithm $\textbf{SampleDom}$ once.
\end{theorem}

\vspace{1\baselineskip}
\begin{proof}
Let $q=\mathrm{poly}(n)$, $k=\mathrm{poly}(\log n)$ and $q\geq (kn)^{2}$,  $h=\lfloor \frac{n}{6\log q}\rfloor $, a security hash $\mathbf{h}: \mathbb{Z}_{+}\rightarrow \{0,1\}^{\ast}$,  $V=k\sqrt{2n\log q}\cdot\log n$.

The simulator $\mathcal{C}$ generates the public key through the following steps:
\begin{itemize}
  \item [(1)] Randomly and uniformly select a matrix $A\in \mathbb{Z}_{q}^{h\times n}$.  We regard $A$ as an instance of the \textbf{SIS} problem.
  \item [(2)] Sample $k$ vectors:  $\bm{\gamma}_1, \ldots, \bm{\gamma}_k \leftarrow \textbf{SampleDome}(1^{n},s)$,  where where $s=V/\sqrt{k}\geq \omega(\sqrt{\log n})$. Let $\bm{\alpha}_{i}=A\cdot\bm{\gamma}_{i}(\mathrm{mod }q)\in \mathbb{Z}_{q}^{h}$ for $i=1,...,k.$  Here, we also require that $\bm{\alpha}_{1},...,\bm{\alpha}_{k}$ are linearly independent. If they are not, repeat this step. According to \cite{33}, the expected value of the number of repetitions will not exceed 2. 

  \item [(3)]   Select a secure collision-resistant hash function  $\mathbf{h}: \mathbb{Z}_{+}=\{0,1\}^{\ast}\rightarrow \{0,1\}^{k}$
  \item [(4)]  Send $pk=(A, \mathbf{h},\bm{\alpha}_{1},...,\bm{\alpha}_{k} )$ to  the adversary $\mathcal{A}$.
\end{itemize}

\vspace{1\baselineskip}
After receiving the public key, the adversary $\mathcal{ A}$ chooses $q_{s}$ messages $\mathbf{x}_{1},...,\mathbf{x}_{q_{s}}\in \mathbb{Z}_{+} $ as signature queries. The simulator $\mathcal{C}$ generates  the signature through the following steps:

\vspace{1\baselineskip}
\begin{itemize}
  \item [(1)]  For the message $\mathbf{x}_{i}$, compute $\mathbf{v}_{i}=\mathbf{h}(\mathbf{x}_{i})$ and $\mathbf{t}_{i}=\sum_{j=1}^{k}\mathbf{v}_{ij}\bm{\gamma}_{j}$.
  \item [(2)] Let $\bm{\sigma}_{i}=\mathbf{t}_{i}$.
  \item [(3)] Output the signatures $\bm{\sigma}_{i}$, for    $1\leq i\leq q_{s}$.
\end{itemize}

Send the signatures $\bm{\sigma}_{1},...,\bm{\sigma}_{q_{s}}$ the adversary $\mathcal{A}$.

\vspace{1\baselineskip}
Next, we prove that the output distribution of the simulator is indistinguishable (within negligible statistical distance) from the output distribution in the real signature scheme.

\vspace{1\baselineskip}
First, in the real signature scheme, the matrix $A$ is generated by the trapdoor generation algorithm \textbf{TrapGen}.  According to Theorem \ref{t3.2}, the matrix $A$ statistically follows a uniform distribution. Secondly, according to Lemma  \ref{l3.4}, $\bm{\alpha}_{1}=A\cdot\bm{\gamma}_{1},...,\bm{\alpha}_{k}=A\cdot\bm{\gamma}_{k}$ also statistically follow a uniform distribution. Therefore, in terms of the generation of the public key, it is statistically indistinguishable from that in the real signature scheme.

\vspace{1\baselineskip}
Secondly, according to Theorem \ref{t3.33}, $\bm{\sigma}_{i}=\mathbf{t}_{i}=\sum_{j=1}^{k}\mathbf{v}_{ij}\bm{\gamma}_{j}$ follows the distribution  $\mathcal{D}_{\mathbb{Z}^{n},V}$. Moreover, since
$$ A\cdot\bm{\sigma}_{i}=A\cdot \mathbf{t}_{i}= A\cdot\sum_{j=1}^{k}\mathbf{v}_{ij}\bm{\gamma}_{j}=\sum_{j=1}^{k}\mathbf{v}_{ij}A\cdot\bm{\gamma}_{j}=\sum_{j=1}^{k}\mathbf{v}_{ij}\bm{\alpha}_{j}=
\bm{\beta}_{i}(\mathrm{mod} q),$$
by Lemma \ref{l3.777}, $\bm{\sigma}_{i}=\mathbf{t}_{i}$ follows the distribution \(\mathcal{D}_{\Lambda, V}\), where $\Lambda_{i}=\Lambda_{q}^{\bm{\beta}_{i}}(A) $. Finally, according to Lemma \ref{l3.2}, we have
$$ \parallel \sum_{j=1}^{k}\mathbf{v}_{ij}\bm{\gamma}_{j} \parallel\leq k\cdot \max_{1\leq j\leq k} \parallel\mathbf{v}_{ij}\parallel\leq k\cdot\frac{V}{\sqrt{k}} \sqrt{n}=V\cdot\sqrt{k\cdot n}.$$
Therefore, the signatures generated by the simulator are also statistically indistinguishable from those of the real scheme.

\vspace{1\baselineskip}
After receiving the signature queries, the adversary $\mathcal{A}$ finally outputs a forged $(\mathbf{x}^{*}, \bm{\sigma}^{*})$, where $\mathbf{x}^{*}\notin \mathrm{span}_{(\parallel)}\{\mathbf{x}_{1},...,\mathbf{x}_{q_{s}}\}$.

\vspace{1\baselineskip}
From verification condition 2), it can be concluded that:
\begin{align*}
A \cdot \bm{\sigma}^{*} (\mathrm{mod} \, q) &= A \cdot [\bm{\sigma}_{1}^{*}, \dots, \bm{\sigma}_{|\mathbf{x}^{*}|}^{*}] (\mathrm{mod} \, q) \\
&= [\bm{\beta}_{1}, \dots, \bm{\beta}_{|\mathbf{x}^{*}|}] \\
&=[\sum_{i=1}^{k}\mathbf{v}_{1j}^{\ast}\cdot\bm{\alpha}_{j}, \dots, \sum_{i=1}^{k}\mathbf{v}_{kj}^{\ast}\cdot\bm{\alpha}_{j}]\\
&= [\sum_{i=1}^{k}\mathbf{v}_{1j}^{\ast}\cdot A\cdot\bm{\gamma}_{j}, \dots, \sum_{i=1}^{k}\mathbf{v}_{kj}^{\ast}\cdot A\cdot\bm{\gamma}_{j}] (\mathrm{mod} \, q).
\end{align*}

Take any arbitrary $i$,  then
$$  A\cdot\bm{\sigma}_{i}^{*} =\sum_{i=1}^{k}\mathbf{v}_{ij}^{\ast}\cdot A\cdot\bm{\gamma}_{j}=A\cdot\sum_{i=1}^{k}\mathbf{v}_{ij}^{\ast}\cdot \bm{\gamma}_{j}(\mathrm{mod}). $$

From verification condition 1), we know that: $\parallel \bm{\sigma}_{i}^{*}\parallel\leq V\cdot\sqrt{k\cdot n}$.  By Lemma \ref{l3.2}, $$\parallel \sum_{i=1}^{k}\mathbf{v}_{ij}^{\ast}\cdot \bm{\gamma}_{j} \parallel\leq k\cdot \frac{V}{\sqrt{k}}\cdot\sqrt{n} = V\cdot \sqrt{k\cdot n}$$
holds with overwhelming probability, so $\parallel \bm{\sigma}_{i}^{*}-\sum_{i=1}^{k}\mathbf{v}_{ij}^{\ast}\cdot \bm{\gamma}_{j}\parallel\leq 2V\cdot \sqrt{k \cdot n}$.

\vspace{1\baselineskip}
If $\bm{\sigma}_{i}^{*}\neq \sum_{i=1}^{k}\mathbf{v}_{ij}^{\ast}\cdot \bm{\gamma}_{j}$,  then the simulator outputs  $\bm{\sigma}_{i}^{*}- \sum_{i=1}^{k}\mathbf{v}_{ij}^{\ast}\cdot \bm{\gamma}_{j}$ as a solution to the $\textbf{SIS}_{q,2V\cdot\sqrt{k\cdot n}}$.
Let the process of outputting $\bm{\sigma}_{i}^{*}-  \sum_{i=1}^{k}\mathbf{v}_{ij}^{\ast}\cdot \bm{\gamma}_{j}$ be denoted as algorithm $\mathcal{B}$. 

\vspace{1\baselineskip}
Next, we discuss the probability that $\bm{\sigma}_{i}^{*}\neq \sum_{i=1}^{k}\mathbf{v}_{ij}^{\ast}\cdot \bm{\gamma}_{j}$.  Since the preimage of the algorithm \textbf{SampleDom} has conditional min-entropy $\omega(\log n)$,
thus
$$  \mathrm{Pr}[\bm{\sigma}_{i}^{*}=\sum_{i=1}^{k}\mathbf{v}_{ij}^{\ast}\cdot \bm{\gamma}_{j} ]\leq 2^{-\omega(\log n )}= \mathrm{negl}(n).$$

Therefore, the simulator can construct an algorithm $\mathcal{B}$ by using the adversary $\mathcal{A}$'s ability that, within time $t+\mathcal{O}(q_{s}T_{\mathcal{SH}}+T_{\mathrm{SampleDom}})$ outputs a solution to the $\textbf{SIS}_{q,2V\cdot\sqrt{k \cdot n}}$ problem with advantage $\epsilon-\mathrm{negl}(n)$.
\end{proof}

\vspace{1\baselineskip}
\textbf{Worst-case connections.}
According to Theorem \ref{t3.1} (\cite{23}), in the situation where \(q \geq \beta\cdot\omega(\sqrt{n\log n})\), the \(\textbf{SIS}_{q,\beta}\) problem is as hard as approximating the SIVP problem in the worst - case scenario with a factor of \(\beta\cdot\tilde{O}(\sqrt{n})\). In the Setup algorithm, the requirement \(q\geq(nk)^2\) serves to ensure that \(q\) has a large enough value for Theorem \ref{t3.1} to be applicable. Even though the exact value of the worst - case approximation factor is determined by the parameter \(k\), it is always a polynomial function of \(n\) regardless of the specific value of \(k\).

\section{A Linear Semigroup-Homomorphic Signature Scheme }

Based on the semigroup-homomorphic signature construction in Section 4, we will construct a secure linear semigroup homomorphic signature. First, we introduce the linear operations on the semigroup.

\vspace{1\baselineskip}
(1) In this section, the message space and signature space we choose are still $((\mathbb{Z}_{+})^{*},\parallel)$ and $((\mathbb{Z}^{n})^{*},\parallel)$ respectively. In addition, we require that the empty  string \( \varnothing \in (\mathbb{Z}_{+})^{*} \), and it satisfies that for any \( \vec{\mathbf{v}} \in (\mathbb{Z}_{+})^{*} \),
\[
\vec{\mathbf{v} }\parallel \varnothing = \varnothing \parallel \vec{\mathbf{v}} = \vec{\mathbf{v}}.
\]
Next, we define the signature of $\varnothing$ as $\aleph$.  It is required that \( \aleph\in (\mathbb{Z}^{n})^{*}  \), and for any \( \bm{\sigma} \in (\mathbb{Z}^{n})^{*} \),
\[
\aleph \parallel \bm{\sigma} = \bm{\sigma} \parallel  \aleph  =\bm{\sigma}.
\]

\vspace{1\baselineskip}
(2) Let \( R = \mathbb{Z}_p = \{ 0, 1, \dots, p-1 \} \), \( c \in \mathbb{Z}_p \), $c\neq0$, and for any \( \vec{\mathbf{v}} \in (\mathbb{Z}_{+})^{*} \), we define the scalar multiplication operation \( c \cdot\mathbf{v} \) as follows:

\[
c \cdot\vec{\mathbf{v}} = \underbrace{\vec{\mathbf{v}} \parallel \cdots \parallel \vec{\mathbf{v}}}_{c}.
\]

If \( c = 0 \), then for any \(\vec{ \mathbf{v}} \in (\mathbb{Z}_{+})^{*} \), we define \( 0 \cdot\vec{\mathbf{v}} = \varnothing \). Therefore, if \( \langle f \rangle = (c_1, \dots, c_k) \), where \( c_i \in \mathbb{Z}_p \), then for any \( \vec{\mathbf{v}}_1, \dots, \vec{\mathbf{v}}_k \in (\mathbb{Z}_{+})^{*} \), we have

\[
f(\vec{\mathbf{v}}_1, \dots, \vec{\mathbf{v}}_k) = c_1 \vec{\mathbf{v}}_1 \parallel \cdots \parallel c_k \vec{\mathbf{v}}_k = \underbrace{\vec{\mathbf{v}}_1 \parallel \cdots \parallel \vec{\mathbf{v}}_1}_{c_1 } \parallel \cdots \parallel \underbrace{\vec{\mathbf{v}}_k \parallel \cdots \parallel \vec{\mathbf{v}}_k}_{c_k }.
\]

(3) Let \( R = \mathbb{Z}_p = \{0, 1, \dots, p-1\} \), \( c \in \mathbb{Z}_p \), $c\neq0$, and for any \( \bm{\sigma} \in (\mathbb{Z}^{n})^{*} \),
we define the scalar multiplication operation \( c \cdot \bm{\sigma} \) as follows:
\[
 c \cdot \bm{\sigma} = \underbrace{\bm{\sigma} \parallel  \cdots \parallel \bm{\sigma}}_{c } = \underbrace{[\bm{\sigma}, \bm{\sigma}, \dots, \bm{\sigma}]}_{c }.
\]
Note: When a signature $\bm{\sigma}$ is written as $[\bm{\sigma}]$, it indicates that it is considered as a matrix at that moment.
If \( c = 0 \), then for any \(\bm{\sigma} \in(\mathbb{Z}^{n})^{*}\), we define \( 0 \cdot \bm{\sigma} = \aleph \). Therefore, if \( \langle f \rangle = (c_1, \dots, c_k) \), where \( c_i \in \mathbb{Z}_p \), then for any \( \bm{\sigma}_1, \dots, \bm{\sigma}_k \in (\mathbb{Z}^{n})^{*}\), we have

\[
f(\bm{\sigma}_1, \dots, \bm{\sigma}_k) = c_1 \cdot\bm{\sigma}_1\parallel \cdots \parallel c_k \cdot\bm{\sigma}_k = [c_1\cdot \bm{\sigma}_1, \cdots, c_k \cdot\bm{\sigma}_k ]=[ \underbrace{\bm{\sigma}_1,\cdots, \bm{\sigma}_1}_{c_1 },\cdots, \underbrace{\bm{\sigma}_k, \cdots, \bm{\sigma}_k}_{c_k } ].
\]

\subsection{Construction}

In the following, we present a formal description of our linear semigroup-homomorphism signature scheme.

\vspace{1\baselineskip}
Our linear semigroup-homomorphic signature scheme  \(\mathcal{LSH}\) = (\textbf{Setup}, \textbf{Sign}, \textbf{Combine},\textbf{ Verify}) works as follows:

\vspace{1\baselineskip}
\textbf{Setup}$(1^{n}, pp)$: The input consists of the security parameter $n$  and the public parameters $pp$. The public parameters $pp$ are defined as follows:
$q=\mathrm{poly}(n)$, $k=\mathrm{poly}(n)$, and $q \geq (kn)^2$, $h = \left\lfloor \frac{n}{6 \log q} \right\rfloor$, $V = k\sqrt{2n \log q} \cdot \log n$.

Furthermore, the public parameters define a message space $\mathcal{M} = ((\mathbb{Z}_{+})^{*}, \parallel)$ and a signature space $\Sigma = ((\mathbb{Z}^{n})^{*},\parallel)$.
 The algorithm generates a pair of public key and secret key as follows:

\begin{enumerate}
  \item [(1)] Compute $(A, T_{A})\longleftarrow \textbf{TrapGen}(q,h ,n)$, where $A\in \mathbb{Z}_{q}^{h\times n}$, $T_{A}$ is a basis for lattice $\Lambda_{q}^{\bot}(A)$.
  \item [(2)] Sample $k$ vectors: $\bm{\alpha}_{1},..., \bm{\alpha}_{k}\stackrel{\$}{\leftarrow} \mathbb{Z}_q^h$. These vectors are linearly independent.
  \item [(3)]Select a secure collision-resistant hash function  $\mathbf{h}: \mathbb{Z}_{+}=\{0,1\}^{\ast}\rightarrow \{0,1\}^{k}$
  \item [(4)] Output the public key $pk=( A,  , \mathbf{h}, \bm{\alpha}_{1},..., \bm{\alpha}_{k} )$ and the secret key $sk=T_{A}$.
\end{enumerate}

\vspace{1\baselineskip}
\textbf{Sign}$(sk, \mathbf{\tau}, \vec{\mathbf{v}})$: Input the secret key $sk=T_{A}$, the tag $\tau\in\{0,1\}^{n}$, the  message $\vec{\mathbf{v}}=\mathbf{v}_{1}\mathbf{v}_{2}\cdot\cdot\cdot \mathbf{v}_{|\vec{\mathbf{v}}|}\in \mathcal{M } $. the signing algorithm proceeds as follows:

\begin{enumerate}
  \item [(1)] Compute $H_{\tau}=\mathrm{diag}\{2\tau_{1}-1,..., 2\tau_{n}-1\}$, where $\tau=(\tau_{1},...,\tau_{n})$.
  \item [(2)] Compute $T_{B^{\tau}}{\leftarrow} \textbf{NewBasis}(A, H_{\tau},T_{A})$, where $T_{B^{\tau}}$ is a basis for $\Lambda_{q}^{\bot}(B^{\tau})$ and  $B^{\tau}=AH_{\tau}^{\top}$. 
  \item [(3)] Compute $\bm{\beta}_{i}=\sum_{j=1}^{k}\mathbf{h}(\mathbf{v}_{i})_{j}\cdot\bm{\alpha}_{j}(\mathrm{mod }q)$, for $1\leq i\leq |\vec{\mathbf{v}}|$.
  \item [(4)] Compute  $\bm{\sigma}_{i}\longleftarrow \textbf{SamplePre}(B^{\tau}, T_{B^{\tau}}, \bm{\beta}_{i}, V)$,  for $1\leq i\leq |\vec{\mathbf{v}}|$. Let $\bm{\sigma}=\bm{\sigma}_{1}\bm{\sigma}_{2}\cdot\cdot\cdot\bm{\sigma}_{|\vec{\mathbf{v}}|}$
  \item [(5)] Output $\bm{\sigma}$  as signature.
\end{enumerate}

\vspace{1\baselineskip}
\textbf{Combine}$(pk,\tau, \{(c_{i}, \bm{\sigma}_{i})_{i=1}^{l}\})$:   Given a public key $pk$, a data-set tag $\tau \in \{0,1\}^n$, and a set of tuples $(c_i, \bm{\sigma}_i)_{i=1}^{l}$, where $c_i \in R = \mathbb{Z}_p$ and $\bm{\sigma}_i \leftarrow \text{Sign}(sk, \tau, \vec{\mathbf{v}}_{i})$, this algorithm outputs a signature $\bm{\sigma} = c_1 \cdot\sigma_1 \parallel\dots \parallel c_l \cdot\bm{\sigma}_l$. (This $\bm{\sigma}$ is intended to be a signature on $c_1 \cdot\vec{\mathbf{v}}_1 \parallel \dots \parallel c_l \cdot\vec{\mathbf{v}}_l$.)

\vspace{1\baselineskip}
 \textbf{Verify}$(pk, \tau, \vec{\mathbf{y}} , \sigma)$.  On input a public key $pk=( A,  , \mathbf{h}, \bm{\alpha}_{1},..., \bm{\alpha}_{k} )$, an data-set tag $\tau\in\{0,1\}^{n}$, a vector $\vec{\mathbf{y}}=\mathbf{y}_{1}\mathbf{y}_{2}\cdot\cdot\cdot\mathbf{y}_{|\vec{\mathbf{y}}|}\in \mathcal{M}$, and  a signature $\bm{\sigma}=\bm{\sigma}_{1}\bm{\sigma}_{2}\cdot\cdot\cdot\bm{\sigma}_{|\vec{\mathbf{y}}|}\in\Sigma$.  The verification process is as follows:
\begin{enumerate}
  \item [(1)]  compute $H_{\tau}=\mathrm{diag}\{2\tau_{1}-1,..., 2\tau_{n}-1\}$ and $B^{\tau}=AH_{\tau}^{\top}$.
  \item [(2)]  compute $\bm{\beta}_{i}=\sum_{j=1}^{k}\mathbf{h}(\mathbf{y}_{i})_{j}\cdot\bm{\alpha}_{j}(\mathrm{mod }q)$, for $i=1,...,|\vec{\mathbf{y}}|$. Let $Y=[\bm{\beta}_{1},...,\bm{\beta}_{|\vec{\mathbf{y}}|}]$.
  \item [(3)] Output 1 if all of the following two conditions are satisfied; otherwise, output 0.
  \begin{enumerate}
    \item [1)]  $\parallel \bm{\sigma} \parallel \leq V\cdot\sqrt{kn}.$
    \item [2)] $B^{\tau}\cdot\bm{\sigma}(\mathrm{mod}q)=Y$. (Note: $B^{\tau}\cdot\bm{\sigma}\equiv B^{\tau}\cdot[\bm{\sigma}_{1},...,\bm{\sigma}_{|\vec{\mathbf{y}}|}]\equiv [B^{\tau}\cdot\bm{\sigma}_{1},...,B^{\tau}\cdot\bm{\sigma}_{|\vec{\mathbf{y}}|}](\mathrm{mod}q)$ )
  \end{enumerate}
\end{enumerate}

\vspace{1\baselineskip}
\textbf{Remark 3:}
Since this signature scheme is essentially built upon the semigroup-homomorphic signature constructed in Section 4, in order to avoid similar situations, we require that this signature scheme can only sign messages $\mathbf{v}\in \mathbb{Z}_{+}$ using the private key.

\subsection{Correctness}

Below, we use a theorem to demonstrate that the above signature system satisfies correctness with overwhelming probability.

\vspace{1\baselineskip}
\begin{theorem}\label{t6.1}
The above linear semigroup-homomorphic signature scheme  satisfies correctness with overwhelming probability.
\end{theorem}

\vspace{1\baselineskip}
\begin{proof}

(1) If  $\bm{\sigma}=[\bm{\sigma}_{1},...,\bm{\sigma}_{|\vec{\mathbf{y}}|}]\longleftarrow \textbf{Sign}(sk, \mathbf{\tau}, \mathbf{y})$, where $\vec{\mathbf{y}}=\mathbf{y}_{1}\mathbf{y}_{2}\cdot\cdot\cdot \mathbf{y}_{|\vec{\mathbf{y}}|}\in \mathcal{M}$, then
$$\bm{\sigma}_{i}\longleftarrow \textbf{SamplePre}(B^{\tau}, T_{B^{\tau}}, \bm{\beta}_{i}, V),$$
where $\bm{\beta}_{i}=\sum_{j=1}^{k}\mathbf{h}(\mathbf{y}_{i})_{j}\cdot\bm{\alpha}_{j}$,  for $i=1,...,|\vec{\mathbf{\mathbf{y}}}|$.

 From Lemma \ref{l3.1} and Theorem \ref{t3.2}, it follows that $\parallel \widetilde{T}_{B^{\tau}}\parallel\leq\parallel \widetilde{HT}_{A}\parallel=\parallel \widetilde{T}_{A}\parallel\leq \mathcal{O}( \sqrt{h\log q})$ with overwhelming probability.
Since $V=k\sqrt{2n\log q}\cdot\log n$, it follows that $\frac{V}{\parallel\widetilde{T}\parallel}\geq k\sqrt{\frac{2n}{h}}\cdot \log n \geq \sqrt{\log n}$, which means $V\geq \parallel \widetilde{T}\parallel\cdot \omega(\sqrt{\log n})$.

Therefore, by Lemma \ref{l3.2}, we have $\parallel \bm{\sigma}_{i}\parallel\leq V\cdot \sqrt{n}\leq V\cdot \sqrt{kn}$ with overwhelming probability. Thus  
$$\parallel\bm{\sigma}\parallel=\max_{1\leq i\leq |\vec{\mathbf{y}}|} \parallel \bm{\sigma}_{i} \parallel\leq V\cdot \sqrt{kn}.$$

Furthermore, since $\bm{\sigma}_{i}\longleftarrow \textbf{SamplePre}(B^{\tau}, T_{B^{\tau}}, \bm{\beta}_{i}, V)$, it follows that $B^{\tau} \cdot\bm{\sigma}_{i}(\mathrm{mod} q)=\bm{\beta}_{i}$. Therefore, $B^{\tau}\cdot\bm{\sigma}(\mathrm{mod}q)=Y$, where $Y=[\bm{\beta}_{1},...,\bm{\beta}_{|\vec{\mathbf{y}}|}]$.

Therefore  $$ \textbf{Verify}(pk, \tau, \vec{\mathbf{y}} , \bm{\sigma})=1. $$

\vspace{1\baselineskip}
(2) If $\bm{\sigma}\longleftarrow \textbf{Combine}(pk,\tau, \{(c_{i}, \bm{\sigma}_{i})_{i=1}^{l}\})$, where $c_{i}\in \mathbb{Z}_{p}$ ,  $\bm{\sigma}_{i}\longleftarrow\textbf{Sign}(sk, \mathbf{\tau}, \vec{\mathbf{v}}_{i}),$ for $1\leq i \leq l$. According to the definition of the algorithm \textbf{Combine},
\[
 \bm{\sigma}= c_1 \cdot\bm{\sigma}_1 \parallel \cdots \parallel c_l \cdot\bm{\sigma}_l = [c_1 \cdot\bm{\sigma}_1, \cdots, c_l\cdot \bm{\sigma}_l ]=[ \underbrace{\bm{\sigma}_1,\cdots, \bm{\sigma}_1}_{c_1 },\cdots, \underbrace{\bm{\sigma}_l, \cdots,\bm{\sigma}_l}_{c_l } ].
\]

It can be seen from 1) that $\parallel \bm{\sigma}_{i}\parallel\leq V\cdot \sqrt{kn}$ with overwhelming probability; thus  $$\parallel\bm{\sigma}\parallel\leq V\cdot \sqrt{kn}. $$

Let $Y_{i}=[ \bm{\beta}_{1}^{i},...,\bm{\beta}_{|\vec{\mathbf{v}}_{i}|}^{i}]$,   then $B^{\tau}\cdot\bm{\sigma}_{i}(\mathrm{mod} q)=Y_{i}$, for $i=1,...,l$.
Therefore, 
\[
 B^{\tau}\cdot\bm{\sigma}(\mathrm{mod} q) =B^{\tau}\cdot[ \underbrace{\bm{\sigma}_1,\cdots, \bm{\sigma}_1}_{c_1 },\cdots, \underbrace{\bm{\sigma}_l, \cdots,\bm{\sigma}_l}_{c_l } ]= [ \underbrace{Y_1,\cdots, Y_1}_{c_1 },\cdots, \underbrace{Y_l, \cdots,Y_l}_{c_l } ].
\]

Let $$\vec{\mathbf{y}}=c_1 \cdot\vec{\mathbf{v}}_1 \parallel \cdots \parallel c_l\cdot \vec{\mathbf{v}}_l = \underbrace{\vec{\mathbf{v}}_1 \parallel \cdots \parallel \vec{\mathbf{v}}_1}_{c_1 } \parallel \cdots \parallel \underbrace{\vec{\mathbf{v}}_l \parallel \cdots \parallel \vec{\mathbf{v}}_l}_{c_l }.$$
Let \( \bm{\beta}_i = \sum_{j=1}^{k}\mathbf{h}(\mathbf{y}_{i})_{j}\cdot\bm{\alpha}_{j} \), for \( i = 1, \dots, |\vec{\mathbf{y}}| \), then

\[
[ \underbrace{Y_1, \dots, Y_1}_{c_1 } ] = [ \bm{\beta}_1, \dots, \bm{\beta}_{c_1 |\vec{\mathbf{v}}|_1} ], \quad \dots, \quad
[ \underbrace{Y_l, \dots, Y_l}_{c_l } ] = [ \beta_{|\vec{\mathbf{y}}| - \sum_{i=1}^{l-1} c_i |\vec{\mathbf{v}}|_i}, \dots, \beta_{\vec{\mathbf{y}}} ].
\]
 Let, $Y=[\bm{\beta}_{1},...,\bm{\beta}_{|\vec{\mathbf{y}}|}]$, thus
 $$    Y= [ \underbrace{Y_1,\cdots, Y_1}_{c_1 },\cdots, \underbrace{Y_l, \cdots,Y_l}_{c_l } ].  $$

 Therefore,  $$  B^{\tau}\cdot\bm{\sigma}(\mathrm{mod }q) = Y .$$

It follows that  $$ \textbf{Verify}(pk, \tau, c_{1}\vec{\mathbf{v}}_{1}\parallel\cdots\parallel c_{l}\vec{\mathbf{v}}_{l} , \textbf{Combine}(pk,\tau, \{(c_{i}, \bm{\sigma}_{i})_{i=1}^{l}\}))=1. $$
\end{proof}

\subsection{Security}

Below, we prove that our signature scheme is existentially unforgeable under adaptive chosen-message attacks with a fixed message range (EUF-CMA-FMR).

\vspace{1\baselineskip}
\begin{theorem}\label{t66}
Let $\mathcal{LHS}$  be the linear semigroup-homomorphic signature described above.  Suppose $q=\mathrm{poly}(n)$, $k=\mathrm{poly}(n)$  and $q\geq (kn)^{2}$, $h=\lfloor \frac{n}{6\log q}\rfloor$,  $V=k\sqrt{2n\log q}\cdot\log n$, $\beta=2V\cdot\sqrt{kn}$ . If the $\textbf{SIS}_{q,\beta}$ problem is hard, then $\mathcal{LSH}$ is is existentially
unforgeable under the adaptive chosen-message attack with a fixed message range (EUF-CMA-FMR).

\vspace{1\baselineskip}
More specifically,  Suppose the adversary $\mathcal{A}$ is allowed to make at most $q_{s}$ chosen-message queries on message subset $\mathbb{Z}_{+}(\subseteq \mathcal{M}=(\mathbb{Z}_{+})^{*})$ and can break the above linear  semigroup-homomorphic signature scheme $\mathcal{LSH}$ with advantage $\epsilon$ within time $t$. Then, the simulator $\mathcal{C}$ can use the adversary's ability to construct an algorithm $\mathcal{B}$ that outputs a solution to the $\mathrm{SIS}_{q,2V\cdot\sqrt{kn}}$ problem with advantage $\epsilon-\mathrm{negl}(n)$ within time $t+\mathcal{O}(q_{s}T_{\mathcal{LSH}}+T_{\mathrm{SampleDom}})$, where $T_{\mathcal{LSH}}$ represents the time required for the simulator to generate a signature, and $T_{\mathrm{SampleDom}}$ represents the time for running the sampling algorithm $\textbf{SampleDom}$ once.
\end{theorem}

\vspace{1\baselineskip}
\begin{proof}

Assume that the adversary $\mathcal{A}$ breaks the signature scheme  $\mathcal{LHS}$  with advantage $\epsilon$ within time $t$. We will now prove that a simulator $\mathcal{C}$ can be constructed, which utilizes the adversary $\mathcal{A}$'s ability to forge signatures to construct a algorithm $\mathcal{B}$ that can output a solution to the $\textbf{SIS}_{q,\beta}$ problem.

Now the simulator $\mathcal{C}$ generates the public key in the following way:

\begin{enumerate}
  \item [(1)] Let the public parameters
 \[
pp = \left\{
\begin{array}{l}
    k = \mathrm{poly}(n), \\
    q = \mathrm{poly}(n) > (kn)^{2}, \\
    h = \left\lfloor \frac{n}{6 \log q} \right\rfloor, \\
    V = k\sqrt{2n \log q} \cdot \log n. 
\end{array}
\right\}.
\]

  \item [(2)]   Randomly and uniformly select a matrix $A\in \mathbb{Z}_{q}^{h\times n}$.  We regard $A$ as an instance of the \textbf{SIS} problem.

  \item [(3)]  Sample $k$ vectors:  $$\bm{\gamma}_1, \ldots, \bm{\gamma}_k \leftarrow \textbf{SampleDome}(1^{n},s),$$  where $s=V/\sqrt{k}\geq \omega(\sqrt{\log n})$. Let $\bm{\alpha}_{i}=A\cdot\bm{\gamma}_{i}(\mathrm{mod }q)\in \mathbb{Z}_{q}^{h}$,  for $i=1,...,k.$

  \item [(4)]   Select a secure collision-resistant hash function  $\mathbf{h}: \mathbb{Z}_{+}=\{0,1\}^{\ast}\rightarrow \{0,1\}^{k}$
  \item [(5)]  Send $pk=(A, \mathbf{h},\bm{\alpha}_{1},...,\bm{\alpha}_{k} )$ to  the adversary $\mathcal{A}$.
\end{enumerate}

\vspace{1\baselineskip}
  After receiving the public key $pk$,  the adversary $\mathcal{A}$ selects a sequence of subsets  $V_{1},...,V_{q_{s}}$  as signature queries, where $V_{i}=span_{\parallel}\{\mathbf{v}_{i1},..., \mathbf{v}_{il}\}$,  and $(\mathbf{v}_{i1},..., \mathbf{v}_{il})\in (\mathbb{Z}_{+})^{k}$, $1\leq i\leq q_{s}$,  where $l $ represents the maximum number of times of the homomorphic operation.. For each $i$ ($i=1,...,q_{s}$), the simulator $\mathcal{C}$ chooses $\tau_{i}$ uniformly from $\{0,1\}^{n}$ and gives the tag $\tau_{i}$ to $\mathcal{A}$. The steps for the simulator to return the signature queries to the adversary $\mathcal{A}$ are as follows:

\begin{enumerate}
  \item [(1)] Compute $H_{\tau}=\mathrm{diag}\{2\tau_{1}-1,..., 2\tau_{n}-1\}$, where $\tau=(\tau_{1},...,\tau_{n})$.
  \item [(2)] Compute $T_{B^{\tau_{i}}}\longleftarrow \textbf{NewBasis}(A, H_{\tau_{i}}, T_{A} )$.  $T_{B^{\tau_{i}}}$ is a basis of $\Lambda_{q}^{\bot}(B^{\tau_{i}})$, where $B^{\tau_{i}}=AH_{\tau_{i}}^{\top}$.
  \item [(3)] Compute 
$$\mathbf{t}_{ij}=H\cdot[\mathbf{h}(\mathbf{v}_{ij})_{1}\cdot\bm{\gamma}_{1}+...+\mathbf{h}(\mathbf{v}_{ij})_{k}\cdot\bm{\gamma}_{k}],$$
 where $\mathbf{h}(\mathbf{v}_{ij})=(\mathbf{h}(\mathbf{v}_{ij})_{1},...,\mathbf{h}(\mathbf{v}_{ij})_{k})$.
  \item [(4)]  Let $\bm{\sigma}_{ij}=\mathbf{t}_{ij}$ ,  for $j=1,...,l$.
\end{enumerate}

The simulator $\mathcal{C}$ output the signature $ \bm{\sigma}_{ij}$, where $i=1,...,q_{s}$ and  $j=1,...,l,$  and sends it to $\mathcal{A} $.

\vspace{1\baselineskip}
Next, we prove that the output distribution of the simulator is indistinguishable (within negligible statistical distance) from the output distribution in the real signature scheme.

By Lemma \ref{l3.7777}, we know that the distribution of $H\cdot[\mathbf{h}(\mathbf{v}_{ij})_{1}\cdot\bm{\gamma}_{1}+...+\mathbf{h}(\mathbf{v}_{ij})_{k}\cdot\bm{\gamma}_{k}]$ is the same as that of $\mathbf{h}(\mathbf{v}_{ij})_{1}\cdot\bm{\gamma}_{1}+...+\mathbf{h}(\mathbf{v}_{ij})_{k}\cdot\bm{\gamma}_{k}$. Therefore, through a discussion entirely similar to that in Theorem \ref{t2}, the output of the simulator and the output of the real scheme are statistically indistinguishable.

\vspace{1\baselineskip}
Finally, the adversary $\mathcal{A}$ generates a valid forgery $ (\tau^{*}, \vec{\mathbf{y}}^{*}, \bm{\sigma}^{*})$,  such that
$$\textbf{Verify}(pk,\tau^{*}, \vec{\mathbf{y}}^{*},\bm{\sigma}^{*})=1,$$
where  $\tau^{*}\in\{0,1\}^{n}$, $\vec{\mathbf{y}}^{*}=\mathbf{y}_{1}^{*}\mathbf{y}_{2}^{*}\cdot\cdot\cdot\mathbf{y}_{|\vec{\mathbf{y}}^{*}|}\in \mathcal{M}$, $\bm{\sigma}^{*}=\bm{\sigma}_{1}^{*}\bm{\sigma}_{2}^{*}\cdot\cdot\cdot\bm{\sigma}_{|\vec{\mathbf{y}}|}^{*}=[\bm{\sigma}_{1}^{*},..., \bm{\sigma}_{|\vec{\mathbf{y}}|}^{*}]\in \Sigma$.

\vspace{1\baselineskip}
Whether it is a \textbf{I-type forgery} or \textbf{II-type forgery}, the following equation holds:

$$  B^{\tau^{*}} \cdot\bm{\sigma} ^{*}(\mathrm{mod} q) = Y^{*},$$

where  $Y^{*}=[\bm{\beta}_{1}^{*},...,\bm{\beta}_{|\vec{\mathbf{y}}|}^{*}]$, $\bm{\beta}_{i}^{*}=\sum_{j=1}^{k}\mathbf{h}(\mathbf{y}_{i}^{*})_{j}\cdot\bm{\alpha}_{j}$, $1\leq i\leq |\vec{\mathbf{y}}^{*}|$. Noting that  $\bm{\alpha}_{j}=A\cdot\bm{\gamma}_{j}(\mathrm{mod }q)$,  we have:

\begin{align*}
    B^{\tau^{*}} \cdot [\bm{\sigma}_{1}^{*}, \dots, \bm{\sigma}_{|\vec{\mathbf{y}}|}^{*}]  (\mathrm{mod} \, q)&= A\cdot H_{\tau^{*}}^{\top} \cdot [\bm{\sigma}_{1}^{*}, \dots, \bm{\sigma}_{|\vec{\mathbf{y}}|}^{*}] \, (\mathrm{mod} \, q) \\
    &= [ A\cdot H_{\tau^{*}}^{\top} \cdot \bm{\sigma}_{1}^{*}, \dots, A\cdot H_{\tau^{*}}^{\top} \cdot \bm{\sigma}_{|\vec{\mathbf{y}}|}^{*} ] \, (\mathrm{mod} \, q) \\
    &= [\bm{\beta}_{1}, \dots, \bm{\beta}_{|\vec{\mathbf{y}}^{*}|}] \\
    &= [\sum_{j=1}^{k}\mathbf{h}(\mathbf{y}_{1}^{*})_{j} \cdot \bm{\alpha}_{j}, \dots, \sum_{j=1}^{k}\mathbf{h}(\mathbf{y}_{|\vec{\mathbf{y}}^{*}|}^{*})_{j} \cdot \bm{\alpha}_{j}] \\
    &= [\sum_{j=1}^{k}\mathbf{h}(\mathbf{y}_{1}^{*})_{j} \cdot A \cdot \bm{\gamma}_{j}, \dots, \sum_{j=1}^{k}\mathbf{h}(\mathbf{y}_{|\vec{\mathbf{y}}^{*}|}^{*})_{j} \cdot A \cdot \bm{\gamma}_{j}]
\end{align*}

Arbitrarily take an index $i\in\{1,...,|\vec{\mathbf{y}}^{*}|\}$, then

$$  A\cdot H_{\tau^{*}}^{\top} \cdot \bm{\sigma}_{i}^{*} (\mathrm{mod }q)=\sum_{j=1}^{k} \mathbf{h}(\mathbf{y}_{i}^{*})_{j} \cdot A \cdot \bm{\gamma}_{j}.$$

Thus,
$$  A\cdot(H_{\tau^{*}}^{\top} \cdot \bm{\sigma}_{i}^{*} -\sum_{j=1}^{k}\mathbf{h}(\mathbf{y}_{i}^{*})_{j} \cdot\bm{\gamma}_{j})(\mathrm{mod }q)= 0$$

From verification condition 1), we know
$$  \parallel H_{\tau^{*}}^{\top}\cdot \bm{\sigma}_{i}^{*}\parallel= \parallel \bm{\sigma}_{i}^{*}\parallel\leq V\sqrt{kn}.$$

By Lemma \ref{l3.2}, $\parallel \sum_{j=1}^{k}\mathbf{h}(\mathbf{y}_{i}^{*})_{j}\cdot\bm{\gamma}_{j} \parallel\leq k\cdot \frac{V}{\sqrt{k}}\cdot\sqrt{n}= V\cdot \sqrt{kn}$ holds with overwhelming probability, so
$$\parallel H_{\tau^{*}}^{\top} \cdot \bm{\sigma}_{i}^{*} -\sum_{j=1}^{k}\mathbf{h}(\mathbf{y}_{i}^{*})_{j}\cdot\bm{\gamma}_{j}\parallel \leq  2V\cdot \sqrt{kn}.$$

If $ H_{\tau^{*}}^{\top} \cdot \bm{\sigma}_{i}^{*} -\sum_{j=1}^{k}\mathbf{h}(\mathbf{y}_{i}^{*})_{j}\cdot\bm{\gamma}_{j}\neq 0$, then the simulator outputs $ H_{\tau^{*}}^{\top} \cdot \bm{\sigma}_{i}^{*} -\sum_{j=1}^{k}\mathbf{h}(\mathbf{y}_{i}^{*})_{j}\cdot\bm{\gamma}_{j} $ as a solution to the $\textbf{SIS}_{q,\beta}$, where $\beta=2V\cdot \sqrt{kn}$.

\vspace{1\baselineskip}
Let the process of outputting  $ H_{\tau^{*}}^{\top} \cdot \bm{\sigma}_{i}^{*} -\sum_{j=1}^{k}\mathbf{h}(\mathbf{y}_{i}^{*})_{j}\cdot\bm{\gamma}_{j}$ be denoted as algorithm $\mathcal{B}$.

\vspace{1\baselineskip}
Finally, we discuss the total probability. Similar to the proof process of Theorem \ref{t2}, we can still obtain that the probability of $ H_{\tau^{*}}^{\top} \cdot \bm{\sigma}_{i}^{*} -\sum_{j=1}^{k}\mathbf{h}(\mathbf{y}_{i}^{*})_{j}\cdot\bm{\gamma}_{j}= 0$ is $\mathrm{negl}(n)$.
Therefore, the simulator can construct an algorithm $\mathcal{B}$ by using the adversary $\mathcal{A}$'s ability that, within time $t+\mathcal{O}(q_{s}T_{\mathcal{LSH}}+T_{\textbf{SampleDom}})$ outputs a solution to the $\textbf{SIS}_{q,\beta}$ problem with advantage $\epsilon-\mathrm{negl}(n)$.
\end{proof}

\subsection{Privacy}

Since our signature scheme only uses the private key to sign messages $\mathbf{v}\in \mathbb{Z}_{+}$, we will prove that the above signature scheme is weak context hiding in this case.

\vspace{1\baselineskip}
\begin{theorem}\label{t67}
If the adversary only queries on the message subset $\mathcal{M}^{\prime}=\mathbb{Z}_{+}$, then the above linearly semigroup-homomorphic signature scheme is weakly context hiding.
\end{theorem}

\vspace{1\baselineskip}
\begin{proof} The challenger runs $\textbf{Setup}(1^{n}, pp)$ to generate a public-private key pair $(pk, sk)$ and sends it to the adversary. Upon receiving the key pair, the adversary outputs $(V_{0}, V_{1}, f_{1}, \dots, f_{s})$, where $V_{0} = \text{span}_{\parallel}\{\mathbf{v}_{01}, \dots, \mathbf{v}_{0k}\}$, $V_{1} = \text{span}_{\parallel}\{\mathbf{v}_{11}, \dots, \mathbf{v}_{1k}\}$, and  $(\mathbf{v}_{i1},..., \mathbf{v}_{il})\in (\mathbb{Z}_{+})^{k}$ for $b = 0, 1$. Additionally, it holds that $f_{i}(\mathbf{v}_{01}, \dots, \mathbf{v}_{0k}) = f_{i}(\mathbf{v}_{11}, \dots, \mathbf{v}_{1k})$ for $i = 1, 2, \dots, s$.

\vspace{1\baselineskip}
In response, the challenger randomly selects a tag $\tau\in \{0,1\}^{n}$ and a random bit $b\in\{0,1\}$.
Let $$\bm{\sigma}_{bi}\leftarrow \textbf{Sign}(sk, \tau, \mathbf{v}_{bi}), i=1,...,k.$$
 Then,  $\bm{\sigma}_{bi}$ is statistically close to $\mathcal{D}_{\Lambda+\bm{\beta}_{bi}, V}$,  where $\Lambda^{}=\Lambda_{q}^{\bot}(B^{\tau})$, $\bm{\beta}_{bi}=\sum_{j=1}^{k} \mathbf{h}(\mathbf{v}_{bi})_{j}\cdot \bm{\alpha}_{j}$.

\vspace{1\baselineskip}
Let \( \langle f \rangle = (c_1, \dots, c_k) \). According to the definition of the algorithm \textbf{Combine}, we have:

\[
\textbf{Combine}(pk, \tau, \{(c_i, \bm{\sigma}_{bi})\}_{i=1}^{k}) \rightarrow \bm{\sigma}_b
= c_1 \bm{\sigma}_{b1} \parallel \dots \parallel c_k \bm{\sigma}_{bk}
= [c_1 \bm{\sigma}_{b1}, \dots, c_k \bm{\sigma}_{bk}]
\]
\[
= [ \underbrace{\bm{\sigma}_{b1}, \dots, \bm{\sigma}_{b1}}_{c_1 }, \dots,
\underbrace{\bm{\sigma}_{bk}, \dots, \bm{\sigma}_{bk}}_{c_k } ]
\]

Note that, according to the definition of the signing algorithm, each column of the above signature $\bm{\sigma}_{b}$ is statistically close to $\mathcal{D}_{\Lambda+\bm{\beta}_{bl},V}$, where $\bm{\beta}_{bl}=\sum_{j=1}^{k}\mathbf{h}(f(\mathbf{v}_{b1},..., \mathbf{v}_{bk})_{l})_{j}\cdot\bm{\alpha}_{j}$, $l=1,...,|f(\mathbf{v}_{b1},...,\mathbf{v}_{bk})|$.

Therefore, the distribution of $\bm{\sigma}_{b}$ depends only on the parameters: $\Lambda$, $\bm{\alpha}_{1},...,\bm{\alpha}_{k}$, $V$, $f$, $f(\mathbf{v}_{b1},..., \mathbf{v}_{bk})$.

Moreover, since $f_{i}(\mathbf{v}_{01},...,\mathbf{v}_{0k})=f_{i}(\mathbf{v}_{11},...,\mathbf{v}_{1k})$, $i=1,2,...,s,$ for any $f_{i}$, $\bm{\sigma}_{0}$ and $\bm{\sigma}_{1}$ they are statistically close to the same distribution.

Consequently, even an unbounded adversary cannot win the privacy game with non-negligible advantage.

\end{proof}

\section{ Conclusions and Open Problems}

In 2002, at the Cryptographers' Track of the RSA Conference, Johnson et al.\cite{3} put forward an open problem: how to construct a secure homomorphic signature on a semigroup instead of on a group.  This paper presents the first homomorphic signature on a semigroup. To prove the security of this scheme, we define a new security model, namely EUF-CMA-FMR (existential unforgeability under adaptive chosen-message attacks within a fixed message range). We prove that the scheme is existentially unforgeable under adaptive chosen-message attacks within a fixed message range (EUF-CMA-FMR), and that it is tightly secure.

Second, we define linear semigroup-homomorphic signatures and the corresponding security model, and extend the semigroup-homomorphic signature constructed in Section 4 to a linear semigroup-homomorphic signature. We prove the correctness, unforgeability, and privacy of this signature scheme.

Finally, there still remain quite a number of unsolved problems here. (1) Regarding the signature length, since the signature length of our signature scheme increases linearly with the increase of the message length, is it possible to construct a semigroup-homomorphic signature with a fixed signature length?  (2) Regarding the message space, is it possible to construct a semigroup-homomorphic signature on other semigroups that do not have inverses? For example, can we construct a set-homomorphic scheme that only permits the union operation? This question is mentioned in \cite{3}. (3) Regarding security, is it possible to construct a semigroup homomorphic signature that satisfies the standard EUF-CMA? (4) Regarding applications, is it possible to construct other linear semigroup-homomorphic signatures, or homomorphic signatures on a semigroup that support two different operations? Furthermore, could we propose a fully homomorphic signature on a semigroup, similar to a fully homomorphic signature scheme?


\end{document}